\documentclass[twocolumn,showpacs,amsfonts,amssymb,amsmath,floats,superscriptaddress,aps]{revtex4-1}

\usepackage{graphicx}

\usepackage{braket}
\def\ie{i.e.\ }

\def\e{\varepsilon}

\def\w{\omega}

\def\k{\vec{k}}
\usepackage{upgreek}
\usepackage{bbold}

\begin{document}
\title{Modeling of gate controlled Kondo effect at carbon point-defects in graphene}

\author{Daniel May}
\affiliation{Theoretische Physik 2, Technische Universit\"at Dortmund,
  44221 Dortmund, Germany}

\author{Po-Wei Lo}
\affiliation{Department of Physics and Center for Theoretical Physics, National Taiwan University, Taipei 10617, Taiwan}
\affiliation{Department of Physics, Cornell University, Ithaca, New York 14853}

\author{Kira Deltenre}
\author{Anika Henke}
\affiliation{Theoretische Physik 2, Technische Universit\"at Dortmund,
  44221 Dortmund, Germany}

\author{Jinhai Mao}
\author{Yuhang Jiang}
\author{Guohong Li}
\author{Eva Y. Andrei}
\affiliation{Rutgers University, Department of Physics and Astronomy, 136 Frelinghuysen Road, Piscataway, NJ 08855 USA}

\author{Guang-Yu Guo}
\affiliation{Department of Physics and Center for Theoretical Physics, National Taiwan University, Taipei 10617, Taiwan}
\affiliation{Physics Division, National Center for Theoretical Sciences, Hsinchu 30013, Taiwan}

\author{Frithjof B. Anders}
\affiliation{Theoretische Physik 2, Technische Universit\"at Dortmund, 44221 Dortmund, Germany}

\date{\today}

\begin{abstract}
We study the magnetic properties in the vicinity of a single carbon defect in a monolayer of graphene.
We include the unbound $\sigma$ orbital and the vacancy induced bound $\pi$ state
in an effective two-orbital single impurity model. The local magnetic moments  are stabilized by
the Coulomb interaction as well as a significant ferromagnetic Hund's rule coupling between
the orbitals predicted by a density functional theory calculation. A hybridization between the orbitals and
the Dirac fermions is generated by the curvature of the graphene sheet in the vicinity of the vacancy.
We present results for the local spectral function calculated using Wilson's numerical renormalization group
approach for a realistic graphene band structure
and find three different regimes depending on the filling, the
controlling chemical potential, and the hybridization strength.
These different regions are characterized by different magnetic properties.
The calculated spectral functions qualitatively agree with recent scanning tunneling spectra on
graphene vacancies.
\end{abstract}

\maketitle

\section{Introduction}
The two $\pi$ orbital sub-bands of a honeycomb lattice in pristine graphene 
realize perfect Dirac
fermions \cite{Novoselov2005,Neto2009} with a linearly vanishing density of
states (DOS) at the Dirac point. As a result, graphene is a semimetal
with a half-filled conduction band \cite{Wallace1947,Neto2009}. While vacancies in metals 
typically act as non-magnetic scattering centres on the 
conduction electrons, it has been shown \cite{Wang2009, Nair2012, Nair2013} 
that removing a carbon atom induces a local magnetic moment that can undergo a Kondo
screening at low temperatures \cite{Chen2011} with substantial Kondo temperatures of order
$T_K\sim 50\,\mathrm{K}$. A tunable Kondo resonance in combination
with the possibility to manipulate the spin degrees of freedom
\cite{Kane2005, Cho2007} may be of fundamental use for practical
graphene based spintronics. 

Resistivity measurements \cite{Chen2011} 
or the magnetic response of graphene \cite{Nair2013} with vacancies
induced by irradiation provide information on ensemble averaged
impurity properties while 
a scanning tunneling microscopy (STM) gives direct access to single vacancy 
qualities. Since the graphene DOS vanishes linearly at the Dirac point such a system
appears to be an ideal system for observing quantum criticality  
\cite{Vojta2010,Fritz2013}. While the Kondo effect is absent for arbitrary coupling
at the Dirac point and particle-hole (PH) symmetry in a linearly vanishing pseudo-gap DOS 
\cite{WithoffFradkin1990,GonzalezBuxtonIngersent1998}, finite doping controlled by an external gate
voltage or strong PH asymmetry can lead to Kondo screening at low enough temperatures.
In addition, the Kondo physics for adatoms such as Co on graphene \cite{Vojta2006,CobaltOnGrapheneLDA2012,Fritz2013,MitchellFritz2013,
RenCoGraphene2014} has drawn some attention in recent years.

While the $3d$ transition metal Co carries an intrinsic magnetic moment, 
the generation of spin moments at carbon vacancies in graphene 
is less clear. Locally, two-orbitals of the three broken sp$^2$ bonds
form a spin singlet bond leaving one radical with free moment \cite{Nair2013}. 
Since these $\sigma$ orbitals are orthogonal to the $\pi$ orbital subsystem responsible
for the Dirac fermions, they do not couple to the conduction band prohibiting the Kondo effect.
Furthermore, the carbon vacancy induces a bound $\pi$ state in the vicinity of the defect
that is energetically located at or close to the Dirac point \cite{Pereira2007,Nanda2012,Cazalilla2012}. 
In a tight-binding formulation for the itinerant $\pi$ electrons, the exact position of the bound state depends on the 
nearest and next nearest neighbor hopping $t$ and $t^\prime$. A vanishing $t^\prime$ results in a symmetrical DOS
and a Dirac Point that coincides with the bound state.
Those bound states are orthogonal to
the itinerant states forming the conduction electron continuum and it is unclear
\cite{Fritz2013} whether these states are responsible for the experimentally observed
magnetic moments \cite{Nair2013}. Due to the local graphene curvature in the vicinity
of the carbon vacancy, the $\sigma$ orbitals and the neighboring $\pi$ orbitals start to
hybridize and the possibility of a Kondo effect emerges. 

In this paper, we investigate the two-orbital model for single vacancies in graphene 
originally proposed by Cazalilla et al.\ \cite{Cazalilla2012} using Wilson's
numerical renormalization group (NRG) approach \cite{Wilson1975,Bulla2008}. 
We calculate local spectra as functions of the gate voltage controlled
chemical potential as well as the hybridization strength which is connected to 
graphene's curvature in the vicinity of the vacancy.
The model comprises one unbound $\sigma$ orbital, the 
locally bound $\pi$ state and the coupling to the remaining $\pi$ continuum.

Mitchell and Fritz used this model \cite{MitchellFritz2013} as starting point
and constructed an effective Kondo model in the local moment regime 
comprising of a spin-1/2 coupled to a logarithmic divergent
effective conduction band density of states \cite{Cazalilla2012, Peres2009}
for a non-interacting $\pi$ subsystem.
The authors include the vacancy
induced bound state in the DOS which is singular at the DP. Thus,
they neglect the Coulomb interaction between $\sigma$ and $\pi$ orbital 
and treat the problem as an effective $s=1/2$ single impurity Kondo
problem.
However, the divergent DOS at the DP is irrelevant for doping
away from charge neutrality.
Since in this paper we are interested in local fluctuations in particular and
treat Coulomb interactions explicitly, we incorporate the bound state in the 
impurity as an effective orbital and use an effective DOS comprising only of
the itinerant graphene $\pi$ states.
 
The scanning tunneling spectra (STS) \cite{Mao2017} indicate
that the system is located closely to local charge degeneracy points 
at which the local orbital occupation is changed by one
indicating that the external gate voltage not only alters the
filling in the graphene layer but also the local charge configuration.
Charge fluctuations become important for
understanding the STS questioning the applicability of the Schrieffer
Wolff transformation in the experimentally relevant parameter space.
Since a density functional
theory (LDA) calculation \cite{Nanda2012} predicted a substantial
ferromagnetic Hund's rule coupling  
$J_H\approx 0.35\,\mathrm{eV}$ between the $\pi$ and the $\sigma$ orbital,
combining the $\pi$ subsystem into a single locally projected density of states (PDOS)
of non-interacting degrees of freedom appears to be an oversimplification.

Previously, single orbital Kondo and Anderson models 
with a graphene type pseudo-gap DOS have been studied using the NRG 
see also the reviews \cite{Vojta2006,Fritz2013}.
Miranda et al.\ \cite{Miranda2014} focused on a disorder-mediated Kondo effect.
Ruiz-Tijerina et al.\ \cite{Tijerina2017} concentrated on the
investigation of a single orbital Anderson impurity model for
different geometries in the context of a dilute ensemble of atomic
impurities in graphene.  

The inclusion of the ferromagnetic coupling \cite{Nanda2012} between
both orbitals, the $\sigma$ orbital and the zero-mode (ZM),  
can favor a local triplet formation in the n doped region
leading to distinct local spin configurations in different parameter spaces, and 
consequently to different STS responses as function of the local graphene curvature 
that match the recent experimental findings \cite{Mao2017} as we will show below. 
We have also explicitly taken into account the hybridization between
the $\sigma$ orbital and the locally bound $\pi$ state as included in
the starting point of Ref.\ \cite{MitchellFritz2013}. This induces a
level repulsion in the two-orbital model as function of the
hybridization strength that turned out to be crucial for the
development of different local charge and spin configurations. 
We investigate the competition between these different charge and spin states
as function of the chemical potential and the hybridization caused by the graphene
curvature. We are able to qualitatively explain the different experimental STS spectra
obtained on different carbon vacancy locations using the same two-orbital model.

The paper is organized as follows. After giving a brief overview on
the experimental motivation and the STS data in Ref.\ \cite{Mao2017} in
Sec.\ \ref{sec:experiment}, we summarize the literature on the
electronic configuration around carbon vacancies in graphene in
Sec.\ \ref{sec:modeling} and introduce a two-orbital model including
our parameterization of the local curvature. 
First, we focus on the charge and spin configurations of the isolated impurity
in Sec.\ \ref{sec:local-scenarios} in order to set the stage for the detailed NRG 
calculation presented in Sec.\ \ref{sec:results}.
Sec.\ \ref{sec:kondo-review} is devoted to the basics of the Kondo
effect in pseudo-gap systems such as graphene.
We begin the result section\ \ref{sec:results} by examining each of the
three relevant hybridization parameter regimes independently -
Sec.\ \ref{sec:weak-regime}, \ref{sec:intermediate-regime}, and
\ref{sec:strong-regime} - before we combine a full parameter scan of
hybridization and chemical potential into a finite temperature diagram
in Sec.\ \ref{sec:phase-diagram}. A detailed analysis of the Kondo
temperatures is presented in Sec.\ \ref{sec:Tk}. We discuss the impact
of parameter changes within the model on the spectral functions as
well as possible extensions of the model in Sec.\
\ref{sec:modified-model} and end with a summary
(Sec.\ \ref{sec:conclusion}) of our findings. 

\section{Experimental Motivation}
\label{sec:experiment}
In a recent STM study \cite{Mao2017}, Mao et al.
have shown that carbon vacancies in graphene exhibit Kondo
physics adjustable by an external gate voltage $V_G$. They identified
two different classes of vacancies clearly distinguishable by their
characteristic scanning tunneling spectra (STS) as a function of
$V_G$. The gate voltage $V_G$ can be directly converted into a
chemical potential $\mu$ separating the system by either p ($\mu<0$)
or n ($\mu>0$) doping. 

The STS of the first class of vacancies show a Kondo resonance for
electron and hole doping with a vanishing $T_K$ close to charge
neutrality as expected from the pseudo-gap DOS in graphene. 
For the second class of vacancies a Kondo peak is only visible in the
hole doped regime and disappears upon approaching charge
neutrality \cite{Mao2017}.

We see the same characteristic behavior in our NRG calculations where
we use the hybridization strength to distinguish both scenarios. Thus,
we will refer to first scenario as the 'intermediate hybridization'
and the second as the 'strong hybridization' regime throughout this
paper. In our calculations, we also identified a third regime, where a Kondo
peak is only found for $\mu>0$. This so-called 'weak hybridization' regime
has been experimentally observed as well \cite{Mao2017}.
Note, that we are dealing with broad crossovers between these regimes
and not sharp phase boundaries. Therefore, we will do without giving
absolute values for the boundaries of the different regimes.

The goal of this paper is to give a physical explanation of all three
regimes using the same two-orbital impurity model and calculate the
spectral function $\rho(\w)$ for those classes by small parameter
changes related to the local curvature of the graphene sheet at the
location of the impurity.

\section{Theory}
\label{sec:theory}

\subsection{Modeling vacancies by a two-orbital Anderson model}
\label{sec:modeling}
\begin{figure}[tbp]
    \includegraphics[width=0.5\textwidth,clip]{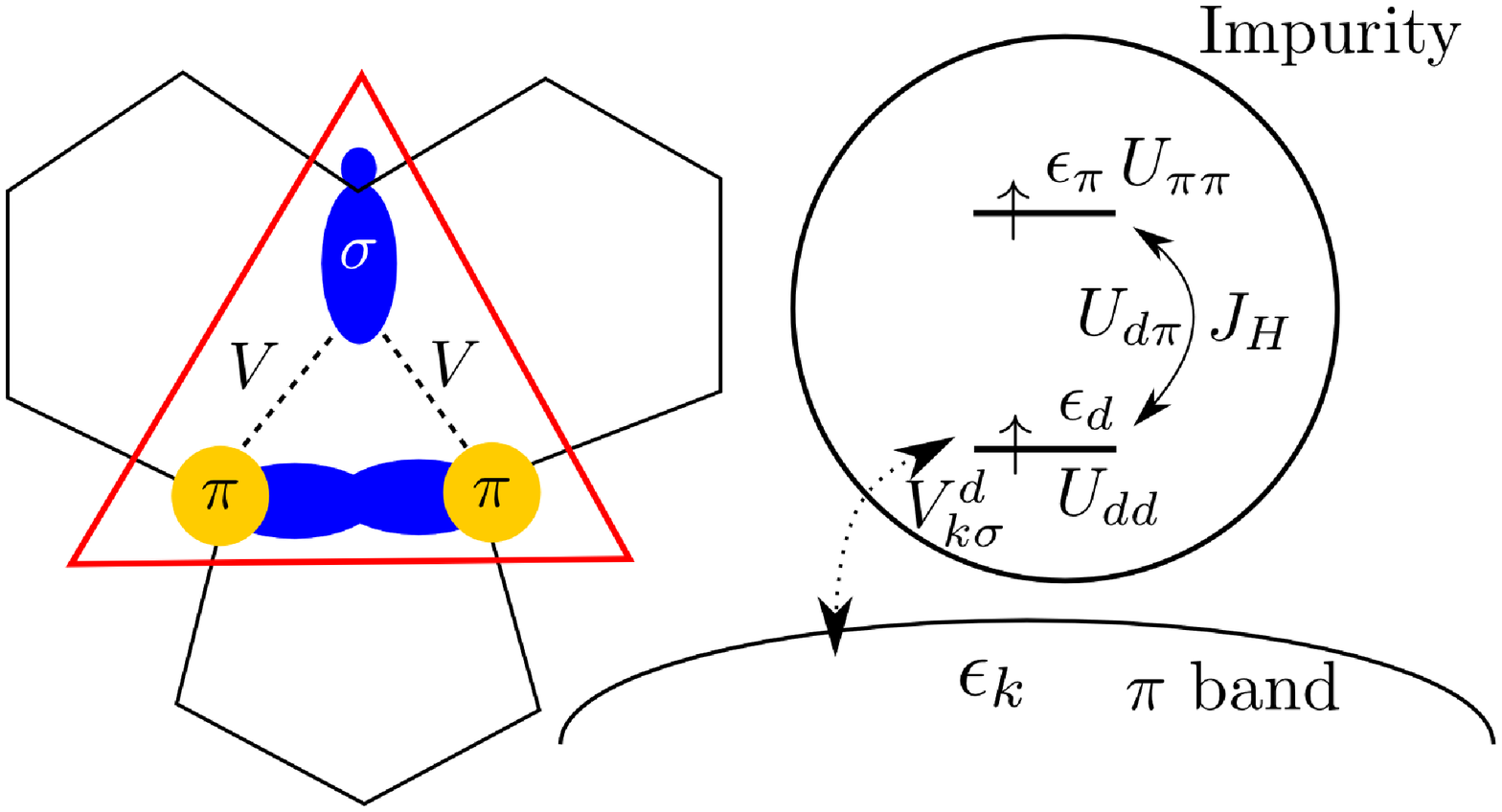}
    \caption{Left: Schematic picture of
      remaining electronic states for a missing carbon atom. The free
      $\sigma$ orbital hybridizes with both adjacent $\pi$ orbitals. 
      The resulting impurity has the characteristic triangle
      shape. Right: Schematic two-orbital model. The $d$ orbital
      represents the free $\sigma$ orbital and is coupled directly to
      the conduction band. The $\pi$ orbital represents the
      vacancy induced zero-mode bound state.
    }
    \label{fig:model-and-lattice}
\end{figure}

Modeling the electronic degrees of freedom in the vicinity of a
graphene vacancy has a long history \cite{Pereira2008}. Removing a carbon atom in a 2D
monolayer of graphene leads to three dangling $\sigma$ orbitals in the
neighboring atoms. Two of them will form a new chemical bond with a
doubly occupied orbital causing a Jahn-Teller like distortion of the
local lattice. Electronically active remains only the third unpaired
$\sigma$ orbital \cite{Cazalilla2012} (see also Fig.~\ref{fig:model-and-lattice}~(left)). Charge neutrality and a weakly
screened local Coulomb interaction causes a free local moment formed
in this chemical radical which can undergo a Kondo screening at low
temperature. 

There is a long-standing debate about the possibility of Kondo
screening of this moment by coupling to the $\pi$ conduction electrons
since the $\pi$ Wannier states are orthogonal to the $\sigma$ orbital
in a flat 2D monolayer of graphene and, therefore, do not hybridize.
This situation however changes in the presence of a local curvature which is naturally
induced by either suspending the sample or by supporting it on a corrugated substrate such
as SiO$_2$. Due to the local curvature, the neighboring $\pi$ orbitals
acquire a finite overlap with the localized $\sigma$ orbital and
start to hybridize with 
a hybridization strength that depends on the local curvature:
the larger the curvature, the stronger the coupling.

In addition, removing a carbon atom from the lattice
also influences the $\pi$ band in the vicinity of the
impurity. Treating this as a single-particle scattering 
problem \cite{Pereira2007} has demonstrated that now the
$\pi$ subsystem comprises itinerant states responsible for the typical
graphene density of states and one additional bound state weakly
localized in the vicinity of the vacancy being  orthogonal to the
itinerant $\pi$ states. This bound state is often referred to as
zero-mode since it is located at the Dirac point in the tight-binding
description \cite{Pereira2007}. Furthermore, density functional theory predicts a
rather significant ferromagnetic coupling between the local
$\pi$ state and the $\sigma$ orbital of $J_H\approx 0.35$eV
\cite{Nanda2012}. 

When setting up an effective model for the vacancy one has to
carefully review the local density of states of the effective
non-interacting conduction band, particularly when hunting for
very subtle changes of many-body wave function due to the Kondo
effect.  

A first guess would be to consider the local density of state (LDOS)
of the neighboring $\pi$ orbitals of the dangling
$\sigma$ orbital. In this LDOS \cite{Pereira2007, Pereira2008} the
bound $\pi$ state causes a singularity at the Dirac point. Using such
LDOS is somehow misleading, since (i) the zero-mode stems for a
$\delta$-distribution that (ii) is associated with an interacting
orbital with a significant Hund's rule coupling \cite{Nanda2012}
which is neglected in the LDOS description.

Alternatively, this bound $\pi$ state could be ignored by
setting up a simplified single-orbital pseudo-gap Anderson model.
Such a model may be sufficient to qualitatively explore the limit
of strong hybridization where only occupation fluctuations between
a singly and a doubly occupied $\sigma$ orbital is relevant -- see
below. However, it fails to capture the physics of the weak
hybridization regime without major modifications and assumptions (e.g. hybridization
dependent Coulomb interaction) which have to be added by hand into the
model. This simplified model also does not address the issue of the
bound state: the resonance resides at $\omega = 0$ yielding a singular
contribution to the DOS that must be properly dealt with. Furthermore,
the ferromagnetic interaction between the $\pi$ bound state and
dangling $\sigma$ orbital is expected to be quite strong and should still be
taken into account explicitly. 

Cazalilla et al.\ \cite{Cazalilla2012} proposed the following
two-orbital Anderson model  
\begin{align}
  \label{eqn:hamil}
  H = H_{\rm loc} + H_{\rm \pi-band} + H_{\rm hyb}
\end{align}
for the description of a carbon vacancy in graphene that will be the
starting point of this paper. The local two-orbital Hamiltonian is
given by 
\begin{align}
  \begin{split}
  \label{eqn:h-loc}
  H_{\rm loc}&= \sum_{\sigma} (\epsilon_{d} n_{d\sigma} + \epsilon_{\pi} n_{\pi\sigma}) + U_{dd} n_{d\uparrow} n_{d\downarrow} \\
             &+ U_{\pi\pi} n_{\pi\uparrow} n_{\pi\downarrow} + U_{d\pi} \sum_{\sigma\sigma'} n_{d\sigma} n_{\pi\sigma} - J_H \vec{S}_\pi  \vec{S}_d \\
             &- J_H (d^\dagger_{d\uparrow} d^\dagger_{d\downarrow} d_{\pi\downarrow} d_{\pi\uparrow} + h.c.) 
           \end{split}
\end{align}
where $\epsilon_{d}$ and $\epsilon_{\pi}$ denotes the single-particle
energy of the orbitals with the corresponding creation operators
$d^\dagger_{\pi(d)\sigma} $ creating an electron in the $\pi$
(respectively $d$) orbital with spin $\sigma$ (we used the subscript $d$
for the $\sigma$ orbital to avoid interference with the spin index)
$U_{\pi\pi}$ and $U_{dd}$ label the intra-orbital Coulomb repulsion
and $U_{d\pi}$ the inter-orbital Coulomb interaction (see Fig.~\ref{fig:model-and-lattice}~(right)). In order to
ensure rotational invariance in the spin space \cite{Oles1983} the
Hund's rule coupling $J_H$ has to be augmented with a pair-hopping
term with the same coupling strength. 

\begin{figure}[tbp]
  \begin{center}
    \includegraphics[width=0.5\textwidth,clip]{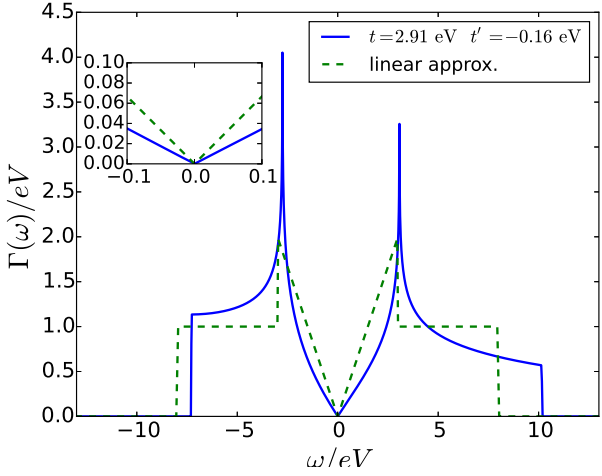}
    \caption{\label{fig:graphene-dos}$\Gamma(\omega)$ for flat graphene using a $t,t'$ tight-binding model with
      $t=2.91\,\mathrm{eV}$ and $t^\prime = -0.16\,\mathrm{eV}$ (solid) in accordance to
      ab-initio calculations \cite{Reich2002, Nanda2009}. We performed a
      summation over the first Brillouin zone and
      evaluated the well-known single-particle dispersion
      Eq.~\eqref{eq:tight-binding-dispersion}. The dashed curve shows a
      simplified approximate $\Gamma(\omega)$. The frequency is
      shifted in both cases such that the Dirac point lies at
      $\omega=0$ for $\mu=0$. 
    }
  \end{center}
\end{figure}

Since we have included the bound state of the $\pi$ system in the
effective impurity, we treat the $\pi$ continuum by a simple
tight-binding model for a bipartite honeycomb lattice as 
\begin{eqnarray}
\label{eq:tight-binding-hamiltonian}
  H_{\rm \pi-band} &=& -t\sum_{<i,j>,\sigma} (a^\dagger_{j\sigma} b_{i\sigma} +b^\dagger_{i\sigma} a_{j\sigma})
  \nonumber \\
                   &&- t'\sum_{\ll i,j\gg ,\sigma}(a^\dagger_{j\sigma} a_{i\sigma} +  b^\dagger_{i\sigma} b_{j\sigma} + h.c.)
\end{eqnarray}
where $\ll i,j\gg $ denotes the summation over a next-nearest neighbor
hopping \cite{Wallace1947}. The operators $a^{(\dagger)}$ and $b^{(\dagger)}$ 
destroy (create) an electron on the respective sublattice A or
B. Fitting of the hopping integrals to DFT  
calculations for the band structure yields the values $t\approx 2.91\,\mathrm{eV}$ 
and $t^\prime\approx -0.16\,\mathrm{eV}$ \cite{Nanda2009} where the inclusion of 
a next-nearest neighbor hopping $t'$ leads to an asymmetrical DOS and a 
shift of the Dirac point. 

Throughout the paper, $\mu=0$ refers to charge neutrality of the system
where the Dirac point coincides with the Fermi energy.
In STM experiments $\mu$ is taken
as the zero of energy and its variation with
the gate induced doping is monitored by the motion of
the Dirac point relative to $\mu$. Likewise, the energy spectrum
is symmetrically discretized around the chemical potential in the NRG. Therefore
the spectral functions are also calculated with respect to $\mu$ as zero of energy.

The tight-binding 
Hamiltonian \eqref{eq:tight-binding-hamiltonian} can be easily diagonalized
which yields two energy bands (cf.\ \cite{Wallace1947, Neto2009})
\begin{align}
\label{eq:tight-binding-dispersion}
	\epsilon_\pm (\vec{k}) &= - t^\prime f(\vec{k}) \pm t\sqrt{3 + f(\vec{k})},\\
	f(\vec{k}) &= 2\cos\left(\sqrt{3} k_y a\right) + 4\cos\left(\frac{\sqrt{3}}{2} k_y a\right)\cos\left(\frac{3}{2} k_x a\right)
\end{align}
where $\tau=\pm$ denotes the band index.

The hybridization energy takes the standard form
\begin{align}
\label{eq:h-hyp}
H_{\rm hyb} = \sum_{\k\sigma\tau} V_{\k}^d( c^\dagger_{\k\sigma\tau} d_{\sigma} +  d^\dagger_{\sigma}c_{\k\sigma\tau} )
\end{align}
where the influence of the $\pi$ band on the impurity dynamics can be
fully encoded into the complex hybridization function
\begin{align}
  \Delta(z) = \sum_{\k, \tau} \frac{|V_{\k}^d|^2}{z-\e_{\tau}(\k)}.
\end{align}
This defines a local Wannier orbital created by $c^\dagger_{0\sigma}$
and an effective hybridization strength $V$ via
\begin{eqnarray}
\label{eq:V-def}
V c^\dagger_{0\sigma} &=& \sum_{\k\sigma\tau} V_{\k}^d  c^\dagger_{\k\sigma\tau}
.
\end{eqnarray}
Taking the imaginary part yields the coupling function
\begin{align}
\label{eq:hybridization}
	\Gamma(\omega) = \pi \sum_{\k, \tau} \vert V_{\k}^d\vert^2 \delta(\omega - \epsilon_\tau(\k)) ,
\end{align}
where $\epsilon_\pm(\k)$ is the one-particle tight-binding solution
Eq.\ \eqref{eq:tight-binding-dispersion}. Since $\{
c^\dagger_{0\sigma}, c_{0\sigma'}\}=\delta_{\sigma\sigma'}$, $V$ is
determined by the equation 
\begin{eqnarray}
V^2 &=& \sum_{\k\sigma} |V_{\k}^d|^2 = \frac{1}{\pi} \int_{-\infty} ^\infty \Gamma(\omega) \mathrm{d}\omega
\end{eqnarray}
that is related to the effective hybridization strength $\Gamma_0$
by the integral
\begin{align}
\label{eq:gamma-norm}
  \int_{-\infty} ^\infty \Gamma(\omega) \mathrm{d}\omega = 2D\Gamma_0 = \pi V^2 ,
\end{align}
where we choose $D =8\,\mathrm{eV}$ for the bandwidth of graphene.

In order to capture the essence of the pseudo-gap density of states, we
also used the parameterization of $\Gamma(\w)$ by the analytical
expression 
($D_{\mathrm{eff}} = 3\,\mathrm{eV}$ pins the van Hove singularities
in the DOS while $D=8\,\mathrm{eV}$ determines the edges)
\begin{align}
\label{eq:approximate}
     \Gamma(\omega) = \Gamma_0 \begin{cases}
      \frac{2\vert\omega\vert}{D_\mathrm{eff}}, & \text{if}\ \vert\omega\vert < D_\mathrm{eff} \\
      0, & \text{if}\ D_\mathrm{eff} \le \vert\omega\vert \le D \\
      1, & \text{otherwise.}
    \end{cases}
\end{align}

The calculations presented below are performed for both
hybridization functions $\Gamma(\w)$, i.e.\ (i) a direct $\k$-space 
summation of the single-particle dispersion  $\epsilon_\tau(\k))$ 
and (ii) the 
approximation \eqref{eq:approximate}. The 
different $\Gamma(\w)$ are depicted in Fig.\ \ref{fig:graphene-dos}.

$\Gamma_0$ is used as a free parameter to adjust the vacancy specific
hybridization introduced above. Thus, it serves as a direct
parameterization of the local curvature in the vicinity of the
impurity. 

\subsection{Local scenarios and choice of parameters}
\label{sec:local-scenarios}

In order to set the stage for the full solution of the correlated
two-orbital impurity problem, we will discuss the different local
configurations in a model which will become relevant for
the explanation of the three different regimes found in the experiments.

In this section, we restrict ourselves to  $H_\mathrm{loc}$ defined in
Eq.\ \eqref{eqn:h-loc}. 
At charge neutrality, the $d$ orbital and the $\pi$ bound state should be singly occupied.
In the p doped region, the $d$ orbital remains
half-filled while the $\pi$ bound state is unoccupied and thus
the total local occupation is typically close to $N_\mathrm{imp}=1$
for $\mu\ll 0$. 

Three distinct local configurations are of particular
interest for positive $\mu$: A second electron can populate either the
$d$ or $\pi$ orbital depending on the local parameters. In addition, a
third electron will occupy the $\pi$ bound state in some parameter
regimes if the additional energy gain $\mu$ is larger than the
differences in the impurity energies. This results in a doublet state
where the $\pi$ orbital is fully occupied and forms a local singlet
whereas the $d$ electron's spin is providing a local moment for a spin-$1/2$ Kondo effect.
 
First, consider the case $N_\mathrm{imp}=2$. The total energy for the
doubly occupied $d$ orbital (singlet state) is given by
\begin{align}
  \label{eq:singlet}
  E_S = 2 \epsilon_d + U_{dd} ,
\end{align}
which competes with the triplet
state where each orbital is singly occupied and whose energy reads
\begin{align}
  \label{eq:triplet}
  E_T =  \epsilon_d + \epsilon_\pi + U_{d\pi} - J_H.
\end{align}
A local singlet ground state formed by two electrons on the $d$ orbital
prohibits the Kondo effect once conduction electrons are coupled
to the orbitals.
On the other hand, the local triplet state still leaves the
possibility for the underscreened Kondo regime with a twofold degenerate
ground state from the $\pi$ orbital spin.

For a total orbital occupation of three
electrons, the energy of the doublet state reads 
\begin{align}
  \label{eq:doublet}
  E_D = \epsilon_d + 2\epsilon_\pi + U_{\pi\pi}  .
\end{align}
Here, the spin of the $d$ electron may also be screened by the
conduction electrons showing Kondo physics. Since the  state $\ket{D}$
contains three electrons and, therefore, its occupation  is
governed by $\Delta E=E_D-3\mu$ in a decoupled impurity.
This state becomes favored with increasing $\mu$ 
over the singlet and the triplet configuration with $N_\mathrm{imp}=2$
and can push the system close to a charge instability.

The local scenario depends sensitively on the parameters of the
local two-orbital Hamiltonian Eq.\ \eqref{eqn:h-loc}. The exchange
interacting $J_H$ is crucial in order to stabilize the local triplet
state. Ab-initio calculations predict a substantial ferromagnetic 
Hund's rules coupling of $J_H\approx 0.35 \,\mathrm{eV}$ \cite{Nanda2012}
between the orthogonal $\pi$ state and $\sigma$ orbital. We propose 
that the difference between the two classes of
impurities detected experimentally and labeled as intermediate 
and strong hybridization regime are related to a hybridization controlled
level crossing between the local singlet state with energy $E_S$ and
the local triplet state with $E_T$.

DFT calculations \cite{Yazyev2007, Padmanabhan2016} and recent
experimental studies \cite{Mao2017} suggest that $U_{dd} =
2\,\mathrm{eV}$ \cite{UestimateMiranda2016}. We will adopt this value but address the
consequence of possible smaller $U$ in Sec. \ref{sec:low-U}.

For the remaining Coulomb matrix elements, we comply with the parameters 
stated in the supplementary material of Ref.\ \cite{Cazalilla2012} where
$U_{d\pi} \approx 0.1 \,\mathrm{eV}$.
In order to stabilize the triplet state in the strong hybridization regime, we slightly
increase the onsite Coulomb repulsion on the $\pi$ orbital, $U_{\pi\pi}
= 0.01\,\mathrm{eV}$. 

Our goal is to switch between the different scenarios by just a
small change in parameters that are solely caused by the
variations in the local curvature of graphene in the vicinity of the different impurities.
We chose the impurity single-particle energies such that the system 
is located close to the local
crossover between singlet and triplet state outlined above. 

Without a microscopic ab-initio theory for each impurity configuration
at hand, we assume that in the experiment only two
parameters, $\Gamma_0$ and $\mu$, are varied: The former by the 
vacancy location, i.\ e.\ the local curvature, the later 
by the external control parameter $V_G$.

The effect of the hybridization between the  
$\pi$ orbitals on the neighboring carbon sites \cite{Kanao2012} and the local $d$ orbital
is twofold: (i) It couples the $d$ orbital to the $\pi$ band continuum and (ii)
generates a hopping term between the bound $\pi$ state and the $d$ orbital.

A single-particle Green function approach \cite{Nanda2012} has been employed 
for the tight-binding model $H_{\rm \pi-band}$ with one carbon vacancy
to determine the fraction $z$ of the local $\pi$ orbital contributing to the  bound $\pi$ state.
The single particle wave function of this bound state
decays only as  $\propto r^{-1}$. Due to the extended nature of the wave function, 
the $z$ factor is relatively small
and has been calculated as $z\approx 0.07$ \cite{Nanda2012}.
Using the hybridization parameter $V$ and diagonalizing
the impurity single-particle matrix
\begin{eqnarray}
  \label{eq:single-particle-orbital-energies}
  H_{\rm loc}^{sp} &=& U^\dagger
  \left(\begin{matrix} \epsilon_d^\prime & 0 \\ 0& \epsilon_\pi^\prime \end{matrix} \right) U
  = \left(\begin{matrix} \epsilon_d & \sqrt{z}V \\ \sqrt{z}V & \epsilon_\pi \end{matrix} \right)
\end{eqnarray}
with the appropriate transformation matrix $U$
determines the new eigenenergies $ \epsilon_d^\prime$ and $ \epsilon_\pi^\prime$ 
of the effective orbitals. Combining an estimate for $\epsilon_d$ and $\epsilon_\pi$
with this hybridization dependent level repulsion induced by the local curvature 
reduces the parameter space to $\Gamma_0$ and $\mu$ as desired.

\begin{figure}[tbp]
   \begin{center}
    \includegraphics[width=0.5\textwidth,clip]{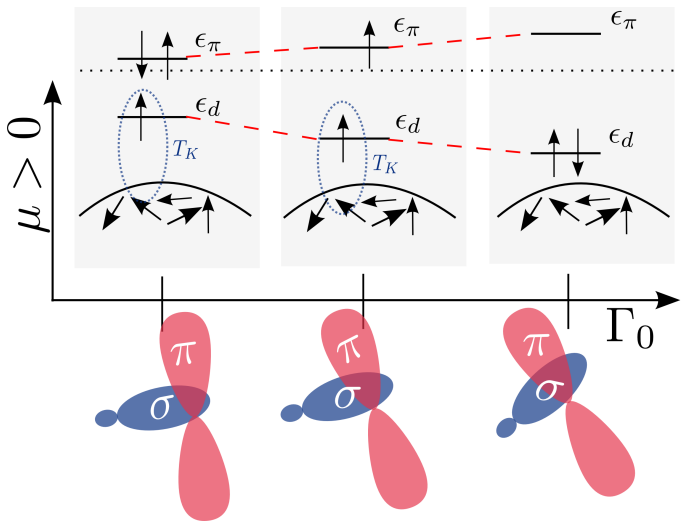}
     \caption{\label{fig:regimes-cartoon}Scenarios for n doping
      depending on the local curvature, i.e. varying $\Gamma_0$ in our
      model. Stronger rippling of the graphene sheet increases the
      overlap between $\sigma$ and $\pi$ orbital resulting in a level
      repulsion. Depending on the strength of $\Gamma_0$ we expect
      three different scenarios as a result of the orbital
      occupation. The ground state energy of the corresponding isolated
      impurity is given by Eq.\eqref{eq:doublet},
      Eq.\eqref{eq:triplet}, and Eq.\eqref{eq:singlet} (left to
      right). Kondo screening is only possible for small and medium  
      hybridization.
    }
  \end{center}
\end{figure}

We also have considered a simplified linear shift interpolating the level energies 
by the equation
\begin{align}
\label{eq:linear-interpolation}
  \epsilon^\prime_{d,\pi}(\Gamma_0) = \epsilon^\prime_{d,\pi}(\Gamma_0^{\mathrm{i}})+ \frac{\epsilon^\prime_{d,\pi}(\Gamma_0^{\mathrm{s}})-\epsilon^\prime_{d,\pi}(\Gamma_0^{\mathrm{i}})}{\Gamma_0^{\mathrm{s}}-\Gamma_0^{\mathrm{i}}}(\Gamma_0-\Gamma_0^{\mathrm{i}}) .
\end{align}
Here, $\Gamma_0^{\mathrm{i,s}}$ stands for the hybridization at
which the system is either in the intermediate or strong hybridization
regime. 
The disadvantage is obvious: There is no microscopic justification
and we need to determine four reference parameters for this interpolation
to arbitrary $\Gamma_0$.

In the following, we omit the prime and implicitly
use the shifted $\Gamma_0$-dependent orbital energies. We also refer  to the extended bound 
$\pi$ state as $\pi$ orbital in the following, since the physical $\pi$ orbitals have been decomposed
in their contribution to $\Gamma(\w)$ and to the bound $\pi$ state.
Since the Coulomb matrix elements
are not determined by an ab-initio approach, we assume that they remain constant
as function of $\Gamma_0$ and are always defined with respect to the final orbital-basis
to keep the modeling simple.

In addition to the previous considerations, the level
positions are subject to a dynamical, $\Gamma_0$-dependent shift
stemming from the interaction with the conduction band continuum
that is explicitly included in our numerical renormalization group approach. As a
consequence, the local estimations for $E_T$ and $E_S$ can only be regarded as
an approximate guideline for the level positions and the charge instability.

For p doping ($\mu <0$), the lower lying $d$ orbital is singly occupied in all
cases eventually resulting in Kondo screening. The exact value of
$T_K$  strongly depends on $\Gamma_0$ and $\mu$. 
Fig.\ \ref{fig:regimes-cartoon} summarises the different
situations for positive $\mu$. For small $\Gamma_0$, the double occupation
of the bound $\pi$ states becomes favorable, leaving a local moment $s=1/2$ in a $N_{\rm imp}=3$
ground state that 
can undergo a Kondo screening at large enough $\mu$. Increasing $\Gamma_0$ will favor
the spin-triplet formation in the local moment regime with $s=1$ that can exhibit an underscreened
Kondo regime at very low temperatures. Increasing $\Gamma_0$ further reduces the
single-particle energy $ \epsilon_d$ enough that the doubly occupied $d$ orbital singlet
state is locally favored, and the possibility of a Kondo effect is excluded. 

\begin{table*}
  \caption{\label{tab:parameters} Parameter sets for the weak,
    intermediate, and strong hybridization regimes for both types of
    DOSs. $\epsilon_{d,\pi}$ represents the newly diagonalized
    orbital.}
  \begin{ruledtabular}
  \begin{tabular}{lllccccccc}
   Regime & DOS & interpolation & $\epsilon_d/\mathrm{eV}$ & $\epsilon_\pi/\mathrm{eV}$ & $U_{dd}/\mathrm{eV}$ & $U_{d\pi}/\mathrm{eV}$ & $U_{\pi\pi}/\mathrm{eV}$ & $J_{H}/\mathrm{eV}$ & $\Gamma_{0}/\mathrm{eV}$\\
  weak         & $t,t^\prime$ & linear     & $-0.93$ & $-0.02$& $2.00$ & $0.10$ & $0.01$ & $0.35$ & $1.10$ \\
               & $t,t^\prime$ & $z$ factor & $-1.21$ & $0.01$ & $2.00$ & $0.10$ & $0.01$ & $0.35$ & $1.00$ \\
  intermediate & approx.     & linear     & $-1.20$ & $0.15$ & $2.00$ & $0.10$ & $0.01$ & $0.30$ & $1.21$ \\
               & $t,t^\prime$ & linear     & $-1.40$ & $0.17$ & $2.00$ & $0.10$ & $0.01$ & $0.35$ & $1.80$ \\
               & $t,t^\prime$ & $z$ factor & $-1.38$ & $0.18$ & $2.00$ & $0.10$ & $0.01$ & $0.35$ & $1.70$ \\
  strong       & approx.     & linear     & $-1.37$ & $0.18$ & $2.00$ & $0.10$ & $0.01$ & $0.30$ & $1.96$ \\
               & $t,t^\prime$ & linear     & $-1.60$ & $0.25$ & $2.00$ & $0.10$ & $0.01$ & $0.35$ & $2.10$ \\
               & $t,t^\prime$ & $z$ factor & $-1.47$ & $0.27$ & $2.00$ & $0.10$ & $0.01$ & $0.35$ & $2.10$ \\
  \end{tabular}
  \end{ruledtabular}
  \end{table*}
  
\subsection{Review of Kondo effect in systems with a pseudo-gap density of states}
\label{sec:kondo-review}  
The Kondo problem has been a major point of interest in sold-state theory
for multiple decades.
It is a prime example of a correlated
many-body problem naturally arising in metallic hosts exhibiting small
magnetic dilution. At sufficient low temperature $T <
T_K$ the antiferromagnetic interaction between the impurity spin and the
electrons' spin of the host gives rise to the formation of a singlet state which screens
the magnetic moment of the impurity. The standard Kondo
Hamiltonian \cite{Wilson1975,Hewson1993} takes the form
\begin{align}
\label{eq:Kondo}
  H = \sum_{\vec{k},\sigma} \epsilon_{\vec{k}} c^\dagger_{\vec{k}\sigma} c_{\vec{k}\sigma} + V_0 \sum_{\vec{k}, \vec{k}^\prime,\sigma} c^\dagger_{\vec{k}\sigma} c_{\vec{k}^\prime\sigma} + J_0 \vec{S} \cdot \vec{s}_0
\end{align}
where $J_0$ is the Kondo coupling between the impurity
spin $\vec{S}$ and the local conduction electron spin density 
$\vec{s}_0$, and $V_0$ accounts for an additional potential scattering term
arising from particle-hole symmetry breaking in the local moment (LM)
regime. 
For a constant metallic DOS $\rho(\omega) \equiv \rho_0$ the Kondo
temperature shows an exponential dependence on $J_0$ \cite{Wilson1975,Hewson1993}
\begin{align}
  T_K = \sqrt{DJ_0} \mathrm{e}^{-1/(J_0\rho_0)}.
\end{align}

For a pseudo-gap density of states, \ie $\rho(\e) \propto |\e|^r$,
Withoff and Fradkin were
the first to point out the existence of a critical coupling
\cite{WithoffFradkin1990} using a perturbative renormalization group
argument. This model exhibits a wide variety of different phases for
$r>0$. These phases are characterized by different fixed points (FP) whose
properties and occurrence depend on the bath exponent $r$ as well as
on particle-hole symmetry or the absence of it. For $0<r<1/2$, there
exists a critical coupling strength $\Gamma_c$ \cite{Chen1995,Ingersent1996,BullaPruschkeHewson1997,GonzalezBuxtonIngersent1998}
governing the transition between a
LM phase for weak coupling and a strong-coupling (SC)
phase for large coupling to the metallic host.

As discussed in section \ref{sec:modeling} and depicted in Fig.\ \ref{fig:graphene-dos},
the case $r=1$ is relevant for graphene. In the following, we briefly summarized the results of
the literature \cite{GonzalezBuxtonIngersent1998, Vojta2010, Fritz2013}.
For $\mu=0$ and PH-symmetry the system will always flow to the stable LM fixed point
independent of the Kondo coupling $J_0$. 
For $\mu\neq 0$, Kondo screening will arise at any
coupling $J_0$ due to the finite density of states and we
expect an exponentially suppressed Kondo scale $\ln(T_K) \propto
-1/\vert\mu\vert^r$ \cite{Vojta2010}.

This picture changes in case of a broken particle-hole symmetry. At $\mu=0$, screening is
possible above a critical coupling $J_c$ and for strong enough asymmetry $V_0 > V_c$.
Thus, there is an unstable quantum critical point separating an unscreened LM and screened ASC
phase. For finite chemical potential and near $J_c$ the system will eventually arrive at a strong coupling 
fixed point but in a different way with regards to particle or hole doping \cite{Vojta2010}.
As a result, close to the quantum critical point (QCP) but for
p doping, the Kondo temperature scales as $T_K = \kappa \vert
\mu\vert$, while for $\mu > 0$ it is proportional to
$\vert\mu\vert^x$, where $x\approx 2.6$ and  
$\kappa$ is a prefactor of order $\mathcal{O}(1)$ that depends only on the
bath exponent $r$ (see Fig.~4 in Ref.~\cite{Vojta2010}).

\subsection{NRG approach to quantum impurity systems}
\label{sec:NRG}
In general, we are interested in the dynamics and the thermodynamics
of an interacting quantum-mechanical many-body
system with an approximate continuous quasi-particle excitation spectrum.
The importance of a wide range of different energy scales, from 
several eV on the scale of the bandwidth to arbitrary small
excitations between degenerate states, gives rise to
infrared divergences in the perturbation theory for the
antiferromagnetic Kondo problem or Anderson impurity models.
One successful approach is the renormalization group (RG) that 
allows for a non-perturbative treatment of all energy scales.
The RG forms the foundation of Wilson's Numerical Renormalization Group 
(NRG) \cite{Bulla2008}
approach which has been first applied to solve the Kondo
problem \cite{Wilson1975, Krishna-murthy1980a, Krishna-murthy1980b}.

The Hamiltonian of such a quantum impurity problem takes the three-part form of
Eq.\ \eqref{eqn:hamil} where the effect of the impurity (band) is completely
encoded in the $H_\mathrm{loc}(H_\mathrm{band})$ term. The bath 
is divided into intervals on a logarithmic mesh, characterized by the 
discretization parameter $\Lambda$, around the chemical potential where 
$\Lambda\rightarrow 1$ reconstructs analytically the original problem. One then proceeds
by mapping this star geometry via a Householder transformation onto a
half-infinite tight-binding chain, the so-called Wilson chain. The
hopping parameter between neighboring sites 
$n$ and $n+1$ decay exponentially $t_n\propto \Lambda^{-n/2}$ owing to the
logarithmic discretization. The problem is now solved in an iterative 
fashion where the starting point is the easily diagonalizable bare
impurity devoid of any bath degree of freedoms. In each iteration
one additional site of the Wilson chain is included, the new
eigenvalue problem is solved numerically, and high-energy excitations are discarded in order
to control the otherwise exponentially growth of the many-body
Fockspace. Each iteration provides access to a smaller temperature
scale due to the exponentially decreasing hopping parameters. The
iterative procedure is then continued until the desired temperature is
reached. In the end, the set of all eigenstates and -energies make up an
approximate solution to the original many-body spectrum up until the
final temperature.

The NRG algorithm is not limited to equilibrium calculations and has
been adopted successfully to non-equilibrium problems
\cite{Anders2005, Anders2006}. The identification of a complete basis
set is the foundation for a sum rule conserving formulation of the
impurity Green's function \cite{Weichselbaum2007, Peters2006} and
eventually for present non-equilibrium extensions. Equilibrium NRG
results provide an almost exact agreement with analytical results if
available and often serve as a benchmark for other methods. For a more
in-depth examination refer to state-of-the-art reviews such as
\cite{Bulla2008}. 
  
\section{Results}
\label{sec:results}
This section starts out with presenting our results for the three different
generic scenarios that are characterized by $\Gamma_0$ before we compile everything into a finite
temperature as well as a $T\rightarrow 0$ regime diagram. 
These diagrams will be presented in Sec.\ \ref{sec:phase-diagram}.
Fig.\ \ref{fig:regimes-cartoon} which summarizes the 
local configurations
for positive chemical potential and three different generic values of $\Gamma_0$
serves as a road map. The scenarios represent the different classes of vacancies 
identified by STM experiments \cite{Mao2017}.
  
We used a discretization parameter $\Lambda=
1.81$ and kept $N_s=2000$ states after each iteration in all our NRG calculations. The
calculations were done for finite temperature $T=4.2\,\mathrm{K}$ 
unless stated otherwise to be comparable to experimental results. 

We employed three different
approximations for $\Gamma(\w)$ and the single-particle 
orbital energies: (i) the
approximate $\Gamma(\w)$ stated in Eq.\ \eqref{eq:approximate} 
and a simple linear interpolation, (ii) the realistic
$\Gamma(\w)$ generated from the $t,t^\prime$ dispersion
and a linear interpolation of $\e_{d/\pi}$, and (iii) the realistic $\Gamma(\w)$ 
as well as a re-diagonalization based on the $z$ factor for the bound
$\pi$ state. 
In all cases, the single particle energies $\epsilon_d$ and $\epsilon_\pi$ 
are functions of the selected $\Gamma_0$.

The parameters used in the NRG calculations are mainly extracted from different 
sources in the literature, as discussed in Sec.\ \ref{sec:local-scenarios} 
and as summarized in table \ref{tab:parameters}. The question whether 
different sets of parameters, especially smaller values for $U_{dd}$, 
results in similar experimental findings is addressed in Sec.\ \ref{sec:low-U}.
In order to reach the desired temperature, we
choose $\overline{\beta} =1.225$ and the number of NRG iterations is
given by $N=34$.

If not stated otherwise, the spectral functions are for the $d$
orbital only, i.e. $\rho_d(\w)$. The effect of the $\pi$ contribution
to the spectrum is discussed in Sec.~\ref{sec:zero-mode}.

\subsection{Weak hybridization regime} 
\label{sec:weak-regime}
\begin{figure}[tbp]
  \begin{center}
    \includegraphics[width=0.5\textwidth,clip]{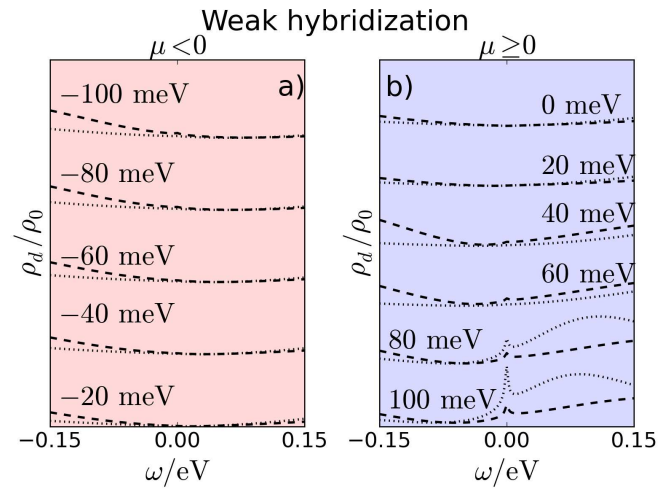}
    \caption{\label{fig:spectral-weak} $\rho_d(\w)$
      for the weak hybridization regime for (a) p doping and (b) n
      doping. The numbers in the plot refer to the chemical
      potential. The results for the realistic $t,t^\prime$
      hybridization function
      combined with simple linear interpolation (dashed) and
      re-diagonalization Eq.\
      \eqref{eq:single-particle-orbital-energies} (dotted) are
      shown. For the approximate hybridization the Kondo effect
        sets in at higher $\mu$ and is not shown for visibility. All
      used parameters are listed in Tab.\ \ref{tab:parameters}. The
      calculations were done for $T=4.2\,\mathrm{K}$. All curves are
      divided by $\rho_0$ being the maximum of all $\rho_d(\w)$.
    }
  \end{center}
\end{figure}

Fig.\ \ref{fig:spectral-weak} shows the calculated
spectral functions $\rho_d(\w)$ \cite{Weichselbaum2007,Peters2006} for the $d$ orbital
for a weak hybridization ($\Gamma_0\approx 1eV$).
The results  are depicted in panel Fig.\ \ref{fig:spectral-weak}(a)
for $\mu<0$ and those for $\mu\ge 0$  in panel Fig.\ \ref{fig:spectral-weak}(b)
and cover a range from  $\mu=-100\,\mathrm{meV}$ to $\mu=100\,\mathrm{meV}$.
For clarity, the spectral functions are shifted by a constant for each
$\mu$ and normalized by $\rho_0$ being the maximum of all spectral functions.
While $\rho_d(\w)$ remains featureless in the p doped regime and reflects the pseudo-gap
DOS of the conduction band, Kondo resonances are visible in the spectra for $\mu>60\,\mathrm{meV}$. 

\begin{figure}[tbp]
  \begin{center}
    \includegraphics[width=0.5\textwidth,clip]{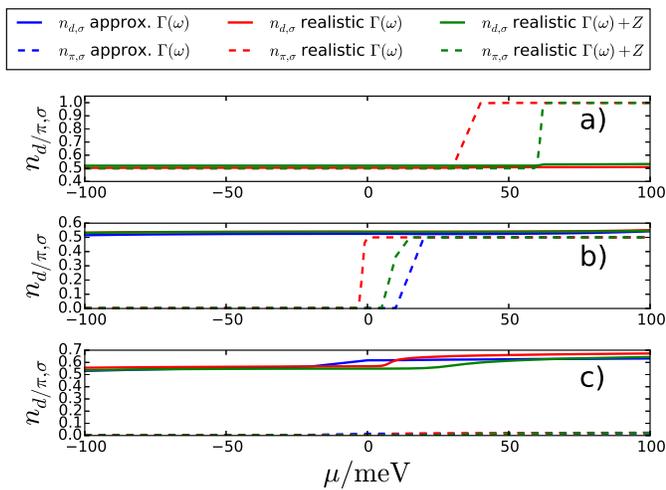}
    \caption{\label{fig:occupation-combined}
    Occupation of the $d$ orbital, $n_{d,\sigma}$, (solid line)
    as well as the ZM represented by the local $\pi$ orbital, $n_{\pi,\sigma}$, (dashed line)
    per spin $\sigma$ at $T=4.2\,\mathrm{K}$
    for (a) weak (b) intermediate (c) strong hybridization regime as calculated
    by the NRG. We used (i) the approximated $\Gamma(\w)$,
    (ii) the realistic $\Gamma(\w)$ obtained from a $t,t^\prime$ TB model
    with a linear interpolation for the orbital evolution, and
    (iii) the realistic $\Gamma(\w)$ in combination with the
    $Z$-factor based orbital splitting according to Eq.~\eqref{eq:single-particle-orbital-energies}.
    }
  \end{center}
\end{figure}

In order to relate the NRG results to the scenarios
presented in Fig.~\ref{fig:regimes-cartoon}, we plot the local orbital
occupations as function of $\mu$ in Fig.~\ref{fig:occupation-combined}(a).
In the weak hybridization regime, the lower $d$
orbital remains half-filled and nearly independent of $\mu$
while the occupation of the $\pi$ orbital 
depends on the chemical potential. 
For $\mu<0$ the $\pi$ orbital is also 
half-filled due to the energy gain by the Hund rule coupling $J_H$
forming a local triplet. Increasing $\mu$ to $\mu>50\,\mathrm{meV}$ adds another electron to
the $\pi$ orbital since the $U_{\pi\pi}$
interaction is weak  and  a local moment with $s=1/2$ remains on the $d$ orbital.
Therefore, we identify an underscreened (undercompensated \cite{CoxZawa98})
Kondo problem for $\mu<0$, while we find a conventional $s=1/2$ pseudo-gap Kondo
problem  for $\mu>0$ since $n_\pi=2$.
The observation of a lower $T_K$ for a underscreened 
$s=1$ Kondo problem in recent NRG calculations (see Ref.\ \cite{Roch2009}) is compatible with our
findings where $T_K\ll T=4.2{\rm K}$ for $\mu<0$.

Additionally, the influence of the chemical potential on the
Kondo temperature is asymmetric even for a fixed $s=1/2$ 
Kondo problem \cite{Vojta2010,Fritz2013} with respect to the 
sign change of $\mu$.

\subsection{Intermediate hybridization regime} 
\label{sec:intermediate-regime}
\begin{figure}[tbp]
  \begin{center}
    \includegraphics[width=0.5\textwidth,clip]{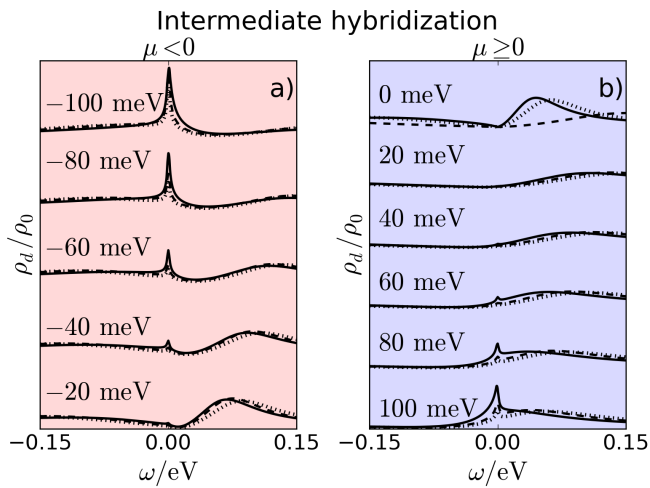}
    \caption{\label{fig:spectral-intermediate}$\rho_d(\w)$ for the
      intermediate hybridization regime for (a) p doping and (b) n
      doping. The numbers in the plot refer to the chemical
      potential. The solid curves represent data obtained with the
      approximate $\Gamma(\w)$, the dashed for the realistic
      $t,t^\prime$, and the dotted for the realistic $\Gamma(\w)$ and
      $Z$ factor. The calculations were done for $T=4.2\,\mathrm{K}$
      and the parameter sets are listed in Tab.\
      \ref{tab:parameters}. All curves are divided by the same
      constant $\rho_0$ which is the maximum of all $\rho_d(\w)$. 
    }
  \end{center}
\end{figure}

Increasing $\Gamma_0$ moves the system into the
intermediate hybridization regime. Here, a Kondo peak  is found in $\rho_d(\w)$ 
for both strong p doping or n doping as clearly depicted in
Fig.\ \ref{fig:spectral-intermediate}. The spectral functions are
normalized as before for clarity.
For small bias voltage, the Kondo effect breaks down since 
$\Gamma(\w)$ vanishes linearly at $\w=0$ suppressing 
screening close to the Dirac point. The Kondo
temperature as function of $\mu$ is extracted using a Goldhaber-Gordon
fit \cite{GoldhaberGordon1998} of the zero
bias conductivity $G(T)$ 
\begin{align}
G(T) = \frac{G_0}{[1+(2^{1/s}-1)(T/T_K)^2]^s},
\end{align}
where $s=0.22$ and $G_0=G(T=0)$. $T_K(\mu)$ shows an exponential
dependence on the chemical potential close to $\vert\mu\vert\rightarrow 0$,
albeit with different slope for p and n doping (not shown).

We have chosen the impurity parameters in the intermediate hybridization regime in
such a way that the occupation of the $\pi$ orbital changes
around $\mu=0$ \cite{Nanda2012} as depicted in Fig.\
\ref{fig:occupation-combined}(b). The $d$ orbital remains roughly
half-filled as function of $\mu$ while the occupancy of the
$\pi$ orbital changes rapidly from zero to single occupation. 
There is only a very small difference
in the results obtained from the approximated $\Gamma(\w)$, Eq.\ \eqref{eq:approximate}, 
and the realistic hybridization function emphasizing the generic character
of the power law approximation of $\Gamma(\w)$. 

Since the local occupancy changes from $N_{\rm imp}=1$ to $N_{\rm imp}=2$,
the underlying Kondo effect is fundamentally different for both types of doping.
For $\mu <0$ the local $d$ spin moment 
is screened by the bath forming a conventional Kondo singlet.
For positive $\mu$, however, the two spins form a local triplet state due
to the strong ferromagnetic Hund's rule coupling $J_H$.
Only a fraction of the resulting $s=1$ moment can be screened by the bath
resulting in an underscreened Kondo effect and a dangling $s=1/2$
moment \cite{CoxZawa98}. 

\subsection{Strong hybridization regime} 
\label{sec:strong-regime}

When $\Gamma_0$ is increased further compared to the 
intermediate regime,
eventually the strong hybridization regime is reached.

\begin{figure}[tbp]
  \begin{center}
    \includegraphics[width=0.5\textwidth,clip]{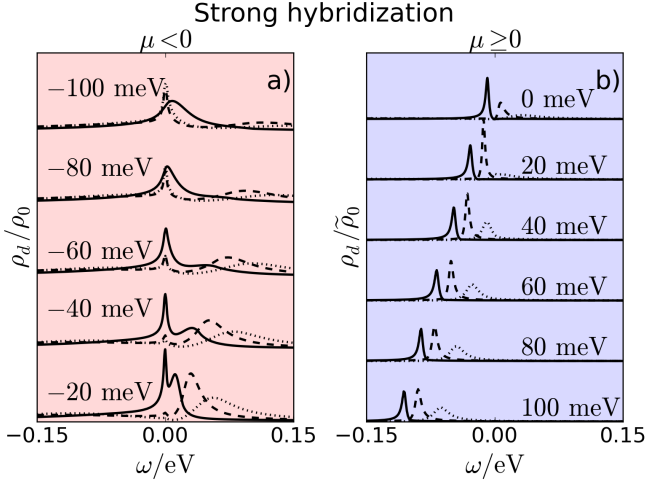}
    \caption{\label{fig:spectral-strong}$\rho_d(\w)$ for the
      strong hybridization regime for (a) p doping and (b) n
      doping. The numbers in the plot refer to the chemical
      potential. The solid curves represent data obtained with the
      approximate $\Gamma(\w)$, the dashed for the realistic
      $t,t^\prime$, and the dotted for the realistic $\Gamma(\w)$ and
      $Z$ factor. The calculations were done for $T=4.2\,\mathrm{K}$
      and the parameter sets are listed in Tab.\
      \ref{tab:parameters}.The spectral functions for n and p doping
      are divided by a different constant each for visibility.
    }
  \end{center}
\end{figure}

For $\mu<0$, it essentially resembles the intermediate
regime. The single occupied $d$ orbital provides an unpaired spin that
is screened to a Kondo singlet, and $\rho_d(\w)$ shows a Kondo resonance
(Fig. \ref{fig:spectral-strong}(a)).
The $\pi$ orbital, however, remains mostly empty for all chemical
potentials considered here and the physics is entirely governed by the $d$
orbital; this regime could also be described by a single orbital Anderson
model \cite{Mao2017}.

Starting around $\mu=0$ and increasing $\mu$, the $d$ orbital gets slightly filled such
that the total occupation approaches $N_{\mathrm{imp}} \approx
1.2$ -- see also Fig.\ \ref{fig:occupation-combined}(c). Therefore, the system
approaches the intermediate valence (IV) FP governed by local charge fluctuations 
\cite{Krishna-murthy1980b,Lo-graphene-2014}. 
A resonance develops in $\rho_d(\w)$ close to
the DP that moves towards lower energies as $\mu$
increases as shown in Fig.\ \ref{fig:spectral-strong}(b). 
This coincides with an increase of the $d$ orbital
occupation. The Kondo resonance for $\mu<0$ evolves continuously
to a charge fluctuation resonance close to $\mu=0$
since the local
$N_{\rm imp}=1$ and $N_{\rm imp}=2$ charge configurations become energetically 
degenerate for $\mu\approx 0$.
While a doubly occupied $d$ orbital is predicted in a purely local picture,
the substantial hybridization favors the itinerancy of the impurity
electrons. The local occupation of the $d$ orbital is shown in Fig.\ \ref{fig:occupation-combined}(c)
and approaches  $n_{d\sigma}\approx 2/3$ indicating the IV FP.

\subsection{Zero-mode peak}
\label{sec:zero-mode}
We only dealt with the spectral functions for the $d$ orbital in
the previous sections. However, the tunneling STS current measured in
the experiments \cite{Mao2017} results from a superposition of three
parts: The $d$ orbital, the $\pi$ orbital as well as the
substrate. 

\begin{figure}[tbp]
  \begin{center}
    \includegraphics[width=0.5\textwidth,clip]{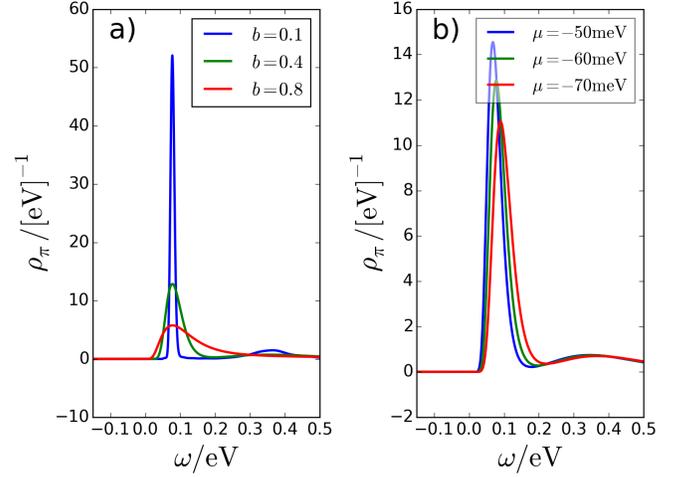}
    \caption{\label{fig:spectral-pi}$\rho_\pi(\w)$ for the
      intermediate hybridization regime for (a) a constant
      $\mu=-60\,\mathrm{meV}$, $\Gamma_0=1.21\,\mathrm{eV}$ and
      varying broadening parameter $b$ and (b) a constant $b=0.4$ and
      varying $\mu$. The approximated $\Gamma(\w)$ and the
      linear interpolation for the level energies were used. All
      parameters are listed in Tab.~\ref{tab:parameters}.
      The calculations were done for $T=4.2\,\mathrm{K}$.
    }
  \end{center}
\end{figure}

The $d$ orbital is responsible for the Kondo physics but
the $\pi$ orbital contributes to the total experimental $\mathrm{d}I/\mathrm{d}V$
curves in form of the zero-mode. The orbital occupation
Fig.~\ref{fig:occupation-combined} gives a rough estimate for the
position of the zero-mode peak relative to the Fermi energy in the
different regimes: a vanishing occupation implies a zero-mode that
rests above $E_F$. We used the same parameters as in the previous sections.
The spectral functions depicted in Fig.~\ref{fig:spectral-pi} 
are calculated by Lehmann representation of the NRG data
\cite{Weichselbaum2007,Peters2006}
using a logarithmic Gaussian broadening $b$ -- 
see the NRG review for details \cite{Bulla2008}.

The $\pi$-spectral functions  in the
intermediate hybridization regime,
$\Gamma_0=1.21\,\mathrm{eV}$, are shown for
varying the broadening parameter $b$ \cite{Bulla2008}
and $\mu=-60\,\mathrm{meV}$ in Fig.~\ref{fig:spectral-pi}(a).
This demonstrates the very sharp nature of the ZM excitation whose
width is artificially broadened by $b$. In addition, a Hund's rule mediated excitation at
slightly higher energies $\omega \approx 0.35 \,\mathrm{eV}$ emerges for small $b$ which
results from the exchange coupling between the $d$ orbital and the $\pi$ orbital.

The zero-mode peak exhibits a simple $\mu$ dependence as shown in
Fig.~\ref{fig:spectral-pi}(b): A higher chemical potential shifts the
peak towards $\omega = 0$. This is the exact same behavior observed in
the experiments (Fig.\ 2(a) in \cite{Mao2017}). 

Since we are mainly interested in the Kondo physics which is only contained in the 
$d$ orbital spectral function we will not discuss the $\pi$ spectral
function in the following. 

\subsection{Mapping out the parameter space}
\label{sec:phase-diagram}

\subsubsection{Regimes at $T=4.2\,\mathrm{K}$}

\begin{figure}[tbp]
  \begin{center}
    \includegraphics[width=0.5\textwidth,clip]{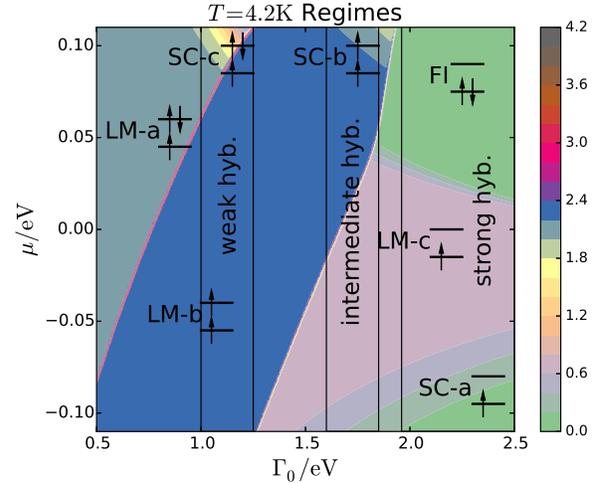}
    \caption{\label{fig:finite-T-phase-diagram}Different
      regimes for finite temperature $T=4.2\,\mathrm{K}$. The
      parameters are listed in Tab.\ \ref{tab:parameters}. We used the
      realistic $\Gamma(\w)$ and the $Z$ factor to determine the level
      positions for arbitrary hybridization strength $\Gamma_0$. For
      the color coding we used the residual entropy times the impurity
      occupation.
}
  \end{center}
\end{figure}

After presenting the $d$ orbital spectral functions for three
different characteristic values for $\Gamma_0$ representing the three different classes of vacancies found
in the STM experiments,
we combine the results into a more elaborate scan of $\Gamma_0$ from
weak to strong hybridization in a single regime diagram where $\mu$
and $\Gamma_0$ are the only free parameters. 

We used the realistic hybridization $\Gamma(\w)$ calculated from the $t,t^\prime$ DOS
and the single-particle orbitals energies 
calculated for a given $\Gamma_0$ via Eq.\ \eqref{eq:single-particle-orbital-energies}
in all NRG calculations in this section.
For each data point $(\Gamma_0,\mu)$, 
we obtained the impurity entropy $S_{\rm imp}$ \cite{Bulla2008}
and total orbital occupation $N_\mathrm{loc}$ from an individual NRG run. The calculations
were performed at a fixed finite temperature $T=4.2\,\mathrm{K}$ to make contact to
the STM experiments. The diagram depicted in Fig.\ \ref{fig:finite-T-phase-diagram}
shows the color-coded product of the residual entropy times the total impurity occupation.

In  Fig.\ \ref{fig:finite-T-phase-diagram} three different areas  
are separated by two naturally appearing lines that define a sharp, but  
temperature broadened crossover region characterized by 
$N_\mathrm{loc}S_{\rm imp}= const$. 
In the left part of the figure, $N_\mathrm{loc}$ rapidly drops from 
three to two, while simultaneously the entropy increases from $\ln(2)$
to $\ln(3)$. The second line occurs where $N_\mathrm{loc}=2\to 1$ and $S_{\rm imp}/k_\mathrm{B}=\ln(3)\to \ln(2)$.

We identify seven different regimes in Fig.\ \ref{fig:finite-T-phase-diagram}. 
Increasing $\Gamma_0$ in the p doped region ($\mu<0$), we
found three local moment and one strong coupling regime: $s=1/2$ (LM-a), $s=1$ (LM-b),
$s=1/2$ (LM-c), and $s=1/2$ (SC-a). 
For positive $\mu$, we find two strong coupling FPs
($s=1$ SC-b and $s=1/2$ SC-c) as well as a
frozen impurity regime (FI).  
Note that the LM regimes mentioned above also
extend partially to $\mu>0$ due to the vanishing $\Gamma(\w)$ at the Dirac point.

For the smallest hybridization $\Gamma_0$ in the range of $0.5{\rm eV}\le \Gamma_0< 0.7{\rm eV}$,
the $\pi$ orbital is doubly occupied for positive $\mu$, and we find a half-filled $d$ orbital.
This regime is labeled as the \mbox{LM-a}  since the
entropy tends towards $S_{\mathrm{imp}}/k_\mathrm{B}\approx\ln(2)$
and the square effective
magnetic moment \cite{Wilson1975} is given by $\mu_{\mathrm{eff}}^2 \approx 0.25$. 
This regime extends into p doping and crosses over into the
SC-c regime upon simultaneous increase of $\mu$ and $\Gamma_0$.

Increasing $\Gamma_0$ at a fixed $\mu$ enhances the level splitting between 
the two local orbitals: the doubly occupied $\pi$ orbital becomes less favorable and the system
crosses over to the $s=1$ (LM-b) regime. The thick red line marks the crossover
from a local occupation of $N_{\rm imp}=3$ to  $N_{\rm imp}=2$
which also distinguishes LM-a and LM-b regimes.
Thus, the entropy in the
LM-b regime tends to $S_{\mathrm{imp}}/k_\mathrm{B} \approx
\ln(3)$ while the effective moment takes the value
$\mu_{\mathrm{eff}}^2 \approx 0.66$. The ground state is a triplet
state formed by both electrons which are coupled due to the strong
ferromagnetic interaction $J_H$. 

Increasing the hybridization further
delocalizes another impurity electron, and the system crosses over to the LM-c
region where only a single $d$ orbital spin is locally present.
The spectral function already shows the onset of a Kondo peak (see
Fig.\ \ref{fig:spectral-intermediate}) even though the entropy
and the effective moment still signifies a LM FP:
$S_{\mathrm{imp}}/k_\mathrm{B} \approx \ln(2)$ and
$\mu_{\mathrm{eff}}^2 \approx 0.25$. 
The $s=1/2$ Kondo temperature is significantly increased above $4.2K$
at very larger $\Gamma_0$ and $\mu \ll 0$, and the crossover from LM-c
to the SC-a regime is observed.
The entropy as well as the square effective magnetic
moment $\mu_{\mathrm{eff}}^2$ tends to zero due to the formation of a
Kondo singlet state in this regime. 

\begin{figure}[tbp]
  \begin{center}
    \includegraphics[width=0.5\textwidth,clip]{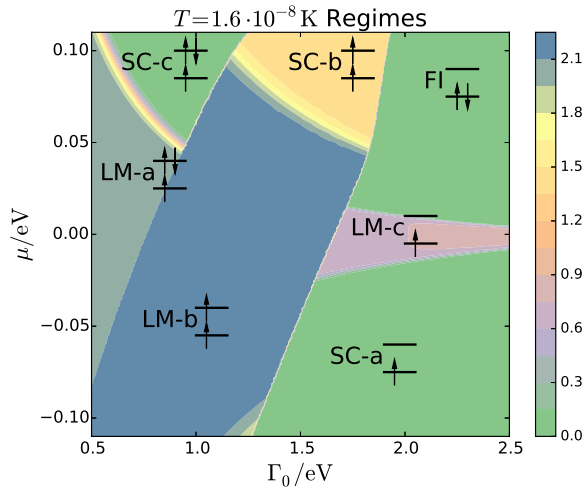}
    \caption{\label{fig:zero-T-phase-diagram}Different
      regimes for temperature $T=1.6\cdot 10^{-8}\,\mathrm{K}$. The
      parameters are listed in Tab.\ \ref{tab:parameters}. We used the
      realistic $\Gamma(\w)$ and the $Z$ factor to determine the level
      positions for arbitrary hybridization strength $\Gamma_0$. For
      the color coding we used the residual entropy times the impurity
      occupation.
    }
  \end{center}
\end{figure}

Now we discuss the diagram for $\mu>0$ starting from 
the upper right area of  Fig.\ \ref{fig:finite-T-phase-diagram}.
For strong enough hybridization and large $\mu$, a doubly occupied
$d$ orbital  and an empty $\pi$ orbital are locally favored.
Although that regime is labeled as  frozen impurity (FI) 
regime, it is closer related the  IV FP  since $N_{\rm imp}=1.2-1.3$.
Interestingly enough, the FI ground state calculated by the NRG
corresponds to a double occupied state while the impurity occupation is closer
to $N_{\mathrm{imp}} \approx 1.3$. Since
the ground state is a local singlet state, the impurity decouples
effectively from the conduction band. The coupling to the conduction band,
however, leads to a gain of kinetic energy and partially delocalizes the $d$ orbital electrons.
The spectral function  shows a pronounced
excitation peak which shifts linearly with the chemical potential
as depicted in Fig.\ \ref{fig:spectral-strong}(b).

Upon the reduction of $\Gamma_0$, the local spin triplet state becomes energetically favorable.
As in the p doped region, we see a broad crossover regime where entropy and
magnetic moment still take their LM fixed point values although an
underdeveloped Kondo peak is already visible -- 
see Fig.\ \ref{fig:spectral-intermediate}(b)
at $\mu=60\,\mathrm{meV}$.
Both local electrons align ferromagnetically, and
the hybridization is sufficient to initiate Kondo screening
of the $d$ electron spin. For sufficiently large $\Gamma(\mu)$,
the underscreened Kondo fixed point (SC-b) is  found with a dangling 
$\pi$ electron spin. The SC-b regime is characterized by  and impurity
entropy of $\ln(2)$ and square effective moment of $0.25$.

At small
$\Gamma_0$ -- upper left corner of  Fig.\ \ref{fig:finite-T-phase-diagram} -- 
we observe another crossover from the LM-a to the
strong coupling regime SC-c. Both regimes are
characterized by a fully occupied $\pi$ orbital. Therefore, we end up
with an effective isolated half-filled $d$ orbital which may now be
screened by the conduction band's electrons when $\Gamma(\mu)$ is increased
with increasing $\mu>0$.

Additionally, we added three areas to the color
contour plot Fig.\ \ref{fig:finite-T-phase-diagram} 
to indicate the three regimes with their distinct spectral evolution
as function of $\mu$: (i) the weak,
(i) intermediate, and
(iii) strong hybridization regime.

\subsubsection{Fixed points $T\to 0$}
Away from the $\mu= 0$ line, all LM regimes  
are associated with unstable RG FPs. The diagram
in  Fig.\ \ref{fig:zero-T-phase-diagram}  is 
calculated
for the same parameters as Fig.\ \ref{fig:finite-T-phase-diagram}
but for $T=\mathcal{O} (10^{-8})\,\mathrm{K}$.
Clearly, the  LM-c regime characterized by a single $d$ orbital spin
is almost completely suppressed and reduced to a narrow region
in the vicinity of $\mu=0$ where
a rapidly vanishing $\Gamma(\w)$ exponentially reduces $T_K$.
This regime has been replaced by the stable SC-a FP
for $\mu<0$ and the FI FP for $\mu>0$. The area of two other stable SC FPs, SC-b and SC-c,
have also increased but the two other LM FP, LM-a and LM-b still cover a large area
of the diagram. Further reduction to $T=10^{-30}K$ -- not shown here
-- reduce these regions in favor of the corresponding SC FPs.

\subsection{Charge crossover from single to double occupation and phase transition at $T\rightarrow
0$}
\label{sec:crossover}
\begin{figure}[tbp]
  \begin{center}
    \includegraphics[width=0.5\textwidth,clip]{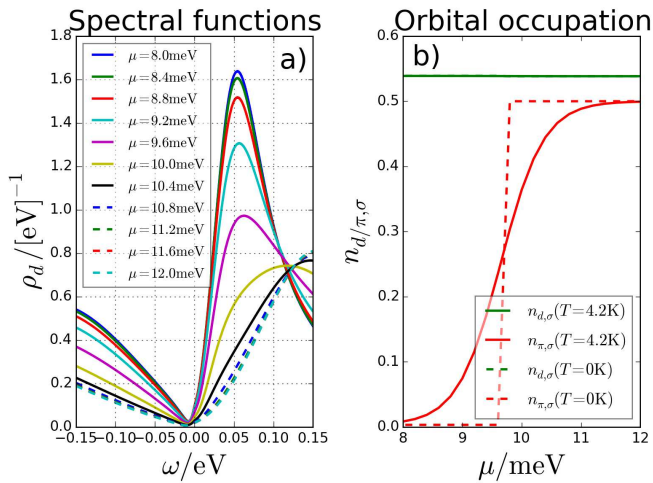}
    \caption{\label{fig:spectral-LM-LM-crossover}(a) Spectral function 
      and (b) orbital occupation for the intermediate regime $\Gamma_0=1.7 \mathrm{eV}$
      obtained for the realistic $\Gamma(\w)$ and $Z$ factor. The spectral
      functions are calculated for finite $T=4.2\,\mathrm{K}$ while
      the occupation is shown for finite $T$ and $T\rightarrow0$ as
      well. The parameters are listed in Tab.\ \ref{tab:parameters}. 
}
  \end{center}
\end{figure}

The question arises whether the crossover from LM-c to LM-b 
in the intermediate regime upon increasing $\mu$  
will develop to a quantum phase transition at $T=0$.
Fig.\ \ref{fig:spectral-LM-LM-crossover} shows the  spectral 
functions, panel (a), as well as the orbital 
occupation, panel (b), in the crossover region.

Already at finite $T=4.2\,\mathrm{K}$ the crossover takes place in a narrow range
between $\mu=8\,\mathrm{meV}$ and $\mu=12\,\mathrm{meV}$ giving rise
to the clearly visible distinction in Fig.\ \ref{fig:finite-T-phase-diagram}. 
The spectral functions for $\mu<\mu_c\approx 9.5{\rm meV}$ are
characterized by a peak at
$\omega \approx 50 \,\mathrm{meV}$ whose spectral weight is shifted towards
slightly higher energies $\omega \approx 115\,\mathrm{meV}$ 
for $\mu>\mu_c$.

The orbital occupation of the impurity illustrates the change from a
crossover to a quantum phase transition at $T=0$. The low lying
$d$ orbital remains at around half-filling regardless of temperature 
and $\mu$. On the other hand, the $\pi$ orbital changes its occupation from empty to
half-filled in a smooth manner at finite $T$. This develops into a
sharp transition for $T\rightarrow 0$ indicating a real phase
transition and a change of ground states.

We see two lines of quantum critical points for our parameters, where
the impurity changes its total occupation by an integral number.

\subsection{The Kondo temperature}
\label{sec:Tk}
In section \ref{sec:phase-diagram} we have shown that the system has not yet
reached its stable fixed points for a large part of the parameter
space at finite $T=4.2\,\mathrm{K}$. The broad crossover regimes show 
qualities which can be ascribed to either a LM or SC phase, such as an
entropy $\propto \ln(2)$ while simultaneously exhibiting an onset of a
Kondo peak. Therefore, estimating the Kondo temperature by using the
full width half maximum (FWHM) of the peak may result in values that
differ significantly from the true low energy scale for the given
parameter sets. In addition, the PH asymmetry and the fitting procedure
does not uniquely determine $T_K$ \cite{AU-PTCDA-monomer}.

We adopted a twofold strategy: (i) we estimated $T_K^{\rm Fano}$
using a Fano line shape for the Kondo peak in the spectral function 
in order to connect to the STS approach for extracting $T_K$ from
spectra experimental obtained over a limited temperature range, 
and (ii) we calculated $T_K$
from the temperature evolution of the zero-bias conductance (ZBC) 
originally proposed by Goldhaber-Gordon \cite{GoldhaberGordon1998} 
in the context to quantum dots that turns out \cite{AU-PTCDA-monomer}
to coincide very accurately with Wilson's definition
\cite{Wilson1975}. However, this method for extracting $T_K$
might be questionable in the case of STS which shows Fano
resonances \cite{SchillerHershfield2000a}.

\begin{figure}[tbp]
  \begin{center}
    \includegraphics[width=0.5\textwidth,clip]{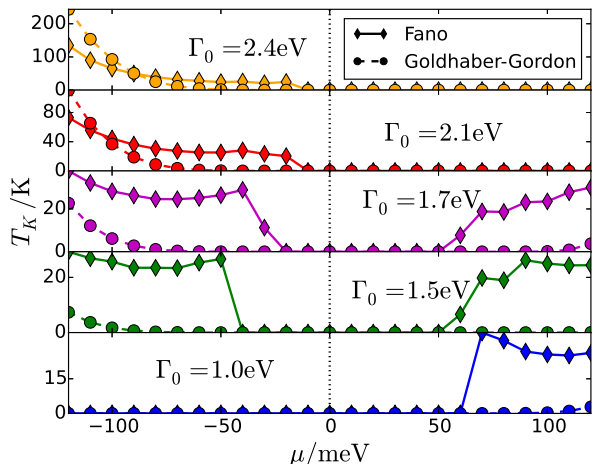}
    \caption{\label{fig:Tk} Kondo temperatures calculated via
      Fano fit extracted from the zero energy peak of $\rho_d(\w)$
      (solid line) and Goldhaber-Gordon fit of the zero bias
      conductivity (dashed line). The Fano fits were carried out at
      finite $T=4.2\,\mathrm{K}$ while the zero bias conductivity was
      calculated at $T=1.3\cdot 10^{-8} \,\mathrm{K}$. The first two
      graphs are from the strong, the next two from the
      intermediate, and the last one from the weak hybridization
      regime. We used the realistic hybridization and Eq.\
      \eqref{eq:single-particle-orbital-energies} for all
      calculations.
}
  \end{center}
\end{figure}

Fig.\ \ref{fig:Tk} shows the calculated $T_K$ values using both
approaches. Note that at 
finite $T=4.2\,\mathrm{K}$ in general the ZBC has not yet reached its
plateau value rendering the Goldhaber-Gordon approach useless, so that
we iterated until the fixed point is reached. For all calculations
we use $\Gamma(\w)$ that stems from the realistic DOS as well as local orbital
energies obtained from $H_{\rm loc}^{sp}$, Eq.\
\eqref{eq:single-particle-orbital-energies}.

The first two curves -- from top to bottom -- are obtained for
hybridization $\Gamma_0 = 2.1,2.4\,\mathrm{eV}$ and can be identified
with the strong hybridization regime, the next two ($\Gamma_0 =
1.5,1.7\,\mathrm{eV}$) show the occurrence of a Kondo peak
characteristic for the intermediate regime, and the last one belongs
to the weak hybridization regime.

Note that the values of $T_K$ determined by the ZBC fit
correspond to the crossover temperature for the entropy change towards
the SC depicted in Fig.\ \ref{fig:finite-T-phase-diagram}. The
agreement between $T_K^{GG}$ extracted from the ZBC fit and true
thermodynamic energy scale $T_K$ characterizing the crossover to the
stable FP in the NRG has been previously observed in the analysis of
STS for Au-PTCDA complexes on Au surface \cite{AU-PTCDA-monomer}.

\subsection{Modification of the model}
\label{sec:modified-model}
All features observed experimentally and reported in Fig.\ 4(a) by Mao
et al.\ \cite{Mao2017} are in qualitative agreement with Fig.\ \ref{fig:Tk}.
However, we also note some differences to the STS data
present in Ref.\ \cite{Mao2017}: 
(i) $T_K^{\rm Fano}$ calculated by the two-orbital model
is about a factor of $1.5-2$ too small and (ii) 
the Kondo resonance in the weak and intermediate
regime sets in at around $\mu\approx 40 \,\mathrm{meV}$ at
the earliest for finite $T=4.2{\rm K}$.
Since the Kondo temperature is exponentially
sensitive to the parameters of the model and the determination
by a FWHM fit in the experiment is not reliable -- see discussion
in Ref.\  \cite{AU-PTCDA-monomer} -- the difference between the experimental and the
theoretical Fano fit $T_K^{\rm Fano}$ is clearly not as significant as
the different onset of the Kondo effect for $\mu>0$.

\subsubsection{Effect of a smaller on-site $U$}
\label{sec:low-U}
The exact value of the intra-orbital Coulomb interaction is not agreed
upon in the literature with predictions spanning roughly one order of
magnitude \cite{Yazyev2007, Nair2013, Wehling2011, Verges2010}.
In the light of a
recent study, we investigate the effect of a
reduced Coulomb matrix element $U_{dd}= 0.5\,\mathrm{eV}$
as proposed by Nair et al.\ \cite{Nair2013}. A smaller Coulomb
interaction enhances local charge
fluctuations and presumably can enhance $T_K$ using the
Schrieffer-Wolff transformation \cite{SchriefferWol66} as a guideline. A larger
$U_{dd}$ on the other hand suppresses those charge fluctuations and, therefore,
requires an even larger, possible non-realistic hybridization $\Gamma_0$. 

Note that $U_{dd} = 0.5\,\mathrm{eV}$ is particularly small in
comparison with the Hund's rule coupling $J_H \approx
0.35\,\mathrm{eV}$ obtained by density functional theory
\cite{Nanda2012} and the inter-orbital
interaction $U_{d\pi} \approx 0.1\,\mathrm{eV}$. In order to guarantee
the occupation necessary for developing a stable local moment, we
slightly increased to $U_{dd} = 0.65\,\mathrm{eV}$. 

The small on-site $U_{dd}$ had to be combined with a value of 
$\epsilon_d = -0.37(-0.40)\,\mathrm{eV}$ in the intermediate (strong)
hybridization regime for the system to form a stable moment in order to find a
Kondo resonance at the chemical potential. Furthermore,
a smaller $U_{dd}$ tended to result in a vacant
impurity ground state abruptly destroying the Kondo effect. 

\begin{figure}[tbp]
  \begin{center}
    \includegraphics[width=0.5\textwidth,clip]{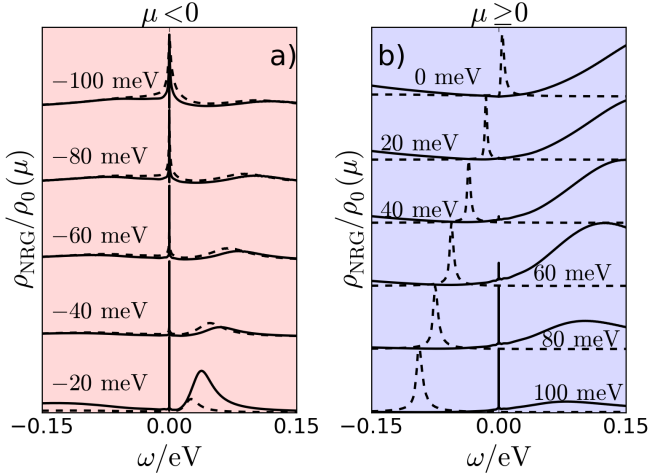}
    \caption{\label{fig:low-u-spectral-functions} $\rho_d(\w)$
      for smaller onsite $U=0.65\,\mathrm{eV}$ and relatively strong Hund's
      coupling $J_H=0.35\,\mathrm{eV}$. The solid (dashed)
      curve represents the intermediate (strong) hybridization regime and the
      impurity parameters are $\epsilon_d=-0.37 (-0.40)\,\mathrm{eV}$,
      $\epsilon_\pi=0.20 (0.23)\,\mathrm{eV}$, and $\Gamma_0 = 0.40 (0.50)\,\mathrm{eV}$. The
      calculations are done for $T=1.6\cdot10^{-5}\,\mathrm{K}$. The
      Kondo peak is clearly pronounced but very narrow indicating a
      small $T_K$.
    }
  \end{center}
\end{figure}

We restricted ourselves to the most approximated
$\Gamma(\w)$ (Eq.\ \eqref{eq:approximate}) for the calculations in order to focus an the
qualitative differences to $U_{dd}=2{\rm eV}$. Fig.\
\ref{fig:low-u-spectral-functions} shows $\rho_d(\w)$ for the modified
parameter set at $T=1.6\cdot10^{-5}\,\mathrm{K}$.
The intermediate (solid lines) and strong hybridization (dashed lines)
regimes are clearly visible with a pronounced Kondo peak. However, the
width of the Kondo resonance is extremely small corresponding to a Kondo
temperature $T_K < 1\,\mathrm{K}$ not comparable to recent
experimental STM data (Ref.\ \cite{Mao2017}). 

Our calculations show that both regimes appear to be
a generic feature of our two-orbital pseudo-gap model.
Counterintuitively, higher Coulomb repulsion $U_{dd} = 2 \,\mathrm{eV}$ yields a
significant higher Kondo temperature and better agreement with STM
experiments but requires an increase of $\Gamma_0$. We could
increase $U_{dd}$ and find adequate parameters for $\Gamma_0$, but
it is not clear whether the corresponding large hybridization matrix element $V$
can be justified by a curvature induced overlap between the local $\sigma$ orbital
and the neighboring tilted $\pi$ orbital, when the $\pi$ orbital matrix elements entering
the band structure are given by $t\approx 2.9{\rm eV}$.

\subsubsection{Possible modifications of the model by neglected interactions}
\label{sec:modified-STM-theory}
In the literature as well as in our calculations, the starting 
point is always a perfect ideal flat graphene sheet. The local modification
by a rippled and curved graphene is only included in (i) the shift of the local
Dirac point and by the finite $\sigma-\pi$ orbital overlap resulting
in a finite $\Gamma_0$. 
We recall that the local curvature is a consequence of an energetically
favored 3D over a purely 2D graphene sheet according to the
Mermin-Wagner theorem and the underlying SiO$_2$ substrate. This will
also modify the phonon modes. Long-wave length Goldstone phonons with
vanishing phonon energies 
will be present in the material but can be neglected for our problem. 
Locally, however, there will be breathing modes of longitudinal vibrations 
of the curved graphene sheet that might expand and contract slightly the distance of
the graphene and the STM tip. Clearly such a local vibration will have a finite frequency
similar to a molecular vibration.  

If we assume the existence of a single local vibrational breathing mode 
it will have two effects: (i) it will modify the Hamiltonian \eqref{eqn:hamil}  
describing the impurity dynamics and (ii) it will modify the tunnel theory employed 
to interpret the STS data.
The definition of $H$ in Eq.\  \eqref{eqn:hamil} has to be augmented by the
two additional terms in $H_\mathrm{ph}$
\begin{eqnarray}
H_\mathrm{ph} &=& \Omega_0 b^\dagger b + \lambda_\mathrm{ph} (b+b^\dagger) H_{\rm hyb}
\end{eqnarray}
where $b$ annihilates a phonon with energy $\Omega_0$.
The first term accounts for free phonons while the second
term result from the expansion of the local hybridization $V$ in 
Eq.\ \eqref{eq:V-def} by
\begin{eqnarray}
V(X) &=& V(1+\lambda_\mathrm{ph} X)
\end{eqnarray}
with the local displacement operator $X=b+b^\dagger$
and the dimensionless electron-phonon coupling strength $\lambda_\mathrm{ph}$. 

Such an additional
term has been investigated within the NRG in the context of 
the quantum transport through a deformable molecular transistor \cite{Cornaglia2005}. 
In the limit of weak electron-phonon coupling, it was analytically shown 
that the Kondo temperature will increase due to an enhancement of the Kondo coupling.
An investigation of the impact onto $T_K$ in the
pseudo-gap two-orbital model as well as an estimation of inelastic tunnel processes
is desirable but beyond the scope of this paper. 

In the context of charge transport through a two-orbital molecule
the influence of an electron-phonon coupling onto an 
additional spin anisotropy term in the spin-1 sector
was investigated \cite{Cornaglia2011, Tijerina2012}. This spin anisotropy 
requires a spin-orbit coupling which is relevant in Co complexes
but weak in carbon. The effect of a small spin anisotropy can be enhanced 
by  a linear spin-lattice coupling in the anti-adiabatic limit resulting in
a reduction of $T_K$. Due to the weak spin-orbit coupling in carbon, we do not
consider this effect to be of relevance  for the spin dynamics of graphene vacancies.
Furthermore, the zero-mode bound state  
is  extended  \cite{Nanda2012}
and can only couple very weakly to
a local phonon.

\subsubsection{Anisotropic hybridization $V$}
\begin{figure}[tbp]
  \begin{center}
    \includegraphics[width=0.5\textwidth,clip]{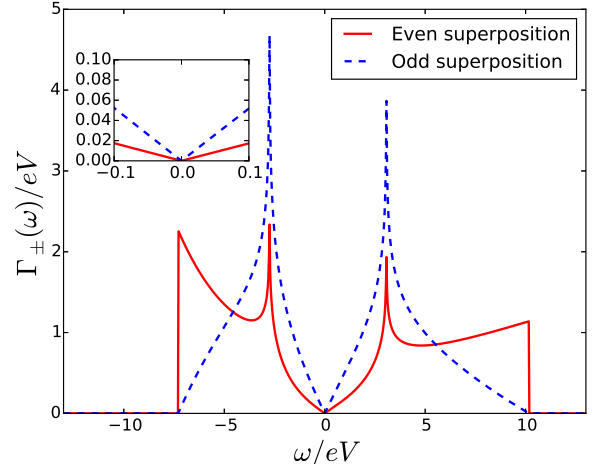}
    \caption{\label{fig:dos-evenodd} $\Gamma_{\pm}(\w)$ stemming from
      the even or odd linear combination in
      Eq.\ \eqref{eq:even-odd-hybridization}. We used a direct
      $\vec{k}$-space summation. For a direct comparison we
      normalized both effective DOSs separately. Both hybridizations
      show the characteristic linear behaviour close to the Dirac Point
      but with different slopes.
    }
  \end{center}
\end{figure}

As indicated in Fig.\ \ref{fig:model-and-lattice}, the dangling
sp$^2$ orbital hybridizes with both $\pi$ orbitals located at the base
of the distorted triangle \cite{Kanao2012}. The hybridization part of the Hamiltonian
can be written as
\begin{align}
\label{eq:local-hyb-II}
  H_\mathrm{Hyb} = \sum_\sigma (V_1 a_{1\sigma}^\dagger d_\sigma + V_2 a_{2\sigma}^\dagger d_\sigma + h.c.)
\end{align}
where $a_{i\sigma}^\dagger$ creates an electron in the $\pi$ orbital
at the nearest-neighbor position $\vec{\delta}_{1,2} = \frac{a}{2}(1,
\pm\sqrt{3})^T$ relative to the vacancy. The localized
$\sigma$ orbital resides at $\vec{\delta}_3 = -a(1,0)^T$.
We consider the general case
where $V_j = \vert V_j \vert \mathrm{e}^{\mathrm{i}\varphi_j}$, i.e. we
allow the hybridization to differ in phase and absolute value. A
point-defect distorts the non-planar geometry of graphene and may lead
to a symmetry breaking hybridization.

Combing the conduction band operators in \eqref{eq:local-hyb-II} to 
the effective operator $Vc_{0\sigma}=  V_1 a_{1\sigma}+ V_2 a_{2\sigma}$
leads to the general hybridization function
\begin{eqnarray}
\tilde \Gamma(\w) &=& \frac{1}{N }\sum_{\k\tau} \left(
|V_1|^2 + |V_2|^2 + 2\Re ( V_1 V_2^* e^{i\vec{k}(\vec{\delta}_1 - \vec{\delta}_2)})
\right)
\nonumber \\
&& \times \delta(\omega - \epsilon_{\vec{k}}^\tau) \ .
\end{eqnarray}
For $\vert V_1 \vert = \vert V_2\vert$, we can follow
the approaches of Refs.\ \cite{Jones-Varma1987, Affleck1995,
  Kanao2012} and combine both $\pi$ orbitals that are situated
opposite the free sp$^2$ (or $\sigma$) orbital to arise at a pair of new orthogonal
operators (even and odd parity). The fermionic commutator determines
the new effective DOS and hybridization function for even ($+$) and
odd ($-$) parity respectively. The corresponding hybridization
functions take the form
\begin{align}
\label{eq:even-odd-hybridization}
\Gamma_\pm(\omega) =
\frac{4|V_1|^2}{N}
\sum_{\vec{k},\tau} 
\delta(\omega - \epsilon_{\vec{k}}^\tau)
\begin{cases}
  \cos^2\left(\frac{\vec{k}(\vec{\delta}_1 - \vec{\delta}_2) + \varphi_2 - \varphi_1}{2}\right)\\
  \sin^2\left(\frac{\vec{k}(\vec{\delta}_1 - \vec{\delta}_2) + \varphi_2 - \varphi_1}{2}\right).
\end{cases}
\end{align}
Here, $\epsilon_{\vec{k}}^\tau$ is the dispersion for the
$\tau$ energy band given by
Eq.\ \eqref{eq:tight-binding-dispersion}. In the case where both $\pi$ orbitals
hybridize equally to the free sp$^2$ orbital, only the even linear
combination couples to the impurity \cite{Kanao2012} while the
odd parity linear combination decouples from the sp$^2$ orbital.
The authors of Ref.\ \cite{Kanao2012} restricted themselves to a linear
dispersion around both Dirac points and a fully symmetric hybridization in which case Eq.\
\eqref{eq:even-odd-hybridization} is analytically solvable for $\vert
V_{\k}^d\vert^2= V^2/N$ and becomes proportional to the zeroth order
Bessel function.

Here, we make use of the full tight-binding dispersion
including second nearest neighbors, Eq.\
\eqref{eq:tight-binding-dispersion}, and evaluate the summation over
the first Brillouin zone explicitly. In the presence of a vacancy the
difference $\vert \vec{\delta}_1 - \vec{\delta}_2\vert$ can be
approximated by the lattice constant $a$ \cite{Kanao2012}.  
Both hybridizations, $\Gamma_\pm(\omega) $,
are shown in
Fig.\ \ref{fig:dos-evenodd}. The characteristic linear dependence close
to the Dirac Point is preserved for both linear combinations. However,
the even linear combination would suppress the Kondo effect
significantly for a fixed $|V_i|$ due to the reduced slope.
In order to retain Kondo
temperatures close to the experimental ones, one could simply increase
$\Gamma_0$ which would eventually lead to questionable high values for
$\Gamma_0$. A relative phase shift of $\delta\varphi=\varphi_2 -
\varphi_1 = \pi$ would result in a swap between even and odd
hybridization and would increase $T_K$. In order for
$\delta\varphi\not = 0$,
either the  configuration belongs to a local odd parity, or the
parity is broken in the vicinity of the vacancy:
both $\pi$ orbitals are slanted differently in
relation to the $\sigma$-plane.

 \begin{figure}[tbp]
   \begin{center}
     \includegraphics[width=0.5\textwidth,clip]{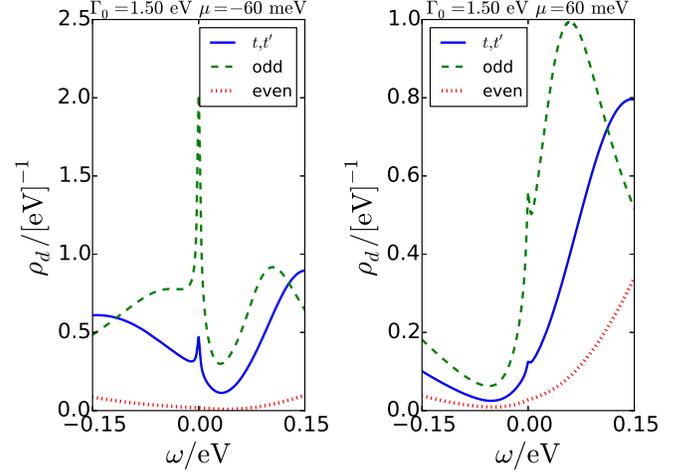}
     \caption{\label{fig:odd-ttprime-comparison} Direct comparison of $\rho_d(\w)$ for 
     (i) $\Gamma(\w)$ calculated with a realistic $t,t^\prime$ DOS (solid), (ii)
     $\Gamma_-(\w)$ as defined in Eq.\
     \eqref{eq:even-odd-hybridization} (dashed), and (iii)
     $\Gamma_+(\w)$ (dotted). All hybridization functions
     are normalized in the same way according to Eq.\
     \eqref{eq:gamma-norm}. We adopted the 
     $t,t^\prime$ parameters of the impurity for the even and odd hybridization function. 
     }
   \end{center}
 \end{figure}

In order to reveal the effect of the different energy dependence of hybridization functions
for the same hybridization strength -- see Eq.\ \eqref{eq:gamma-norm}  --- 
we present a direct comparison of a spectral functions from the
intermediate regime 
at $T=4.2\,\mathrm{K}$ for $\Gamma(\w)$ based on a realistic $t,t^\prime$ DOS 
used throughout this paper and the even/odd hybridization as defined in Eq.\ \eqref{eq:even-odd-hybridization}
in Fig.\ \ref{fig:odd-ttprime-comparison} using identical local impurity parameters.
While the Kondo temperature is strongly suppressed for the even $\Gamma_+(\w)$, we
observe an enhanced Kondo temperature for $\Gamma_-(\w)$ illustrating the effect of
the significantly large slope of the $\Gamma(\w)\propto |\w|$ region
in the vicinity of the Dirac point.
Note that, in order to compare the consequence of the changed hybridization throughout, the parameters of the impurity 
ought to be adjusted.

In a flat graphene sheet with parity conservation at 
the location of the $\sigma$ orbital, the even hybridization function $\Gamma_+(\w)$
should be relevant \cite{Kanao2012}.
Due to the orthogonality of the Wannier orbitals, however, a
hybridization is only generated for curved graphene in the vicinity of
the carbon vacancy. This locally distorted environment breaks parity in
general. The matrix elements $V_i$ will be different in modulus and
phase, and the most general hybridization function is given by
\begin{eqnarray}
\tilde \Gamma(\w) &=& \frac{|\bar V|^2}{N }\sum_{\k\tau} \delta(\omega - \epsilon_{\vec{k}}^\tau)
\nonumber \\
&& \times
\left(1 
 + \frac{2|V_1||V_2|}{|V|^2} 
  \cos\left(\vec{k}(\vec{\delta}_1 - \vec{\delta}_2) + \delta \varphi\right)
\right) 
\end{eqnarray}
defining $|V_1|^2 + |V_2|^2= |\bar V|^2$. If $\vec{k}_F(\vec{\delta}_1 - \vec{\delta}_2) + \delta \varphi\approx 0$,
$\tilde\Gamma(\w) \propto \Gamma(\w)$  close to the Dirac point.
Since the parity breaking should 
be taken into account  and a detailed microscopic theory of the specific single-particle properties
in the vicinity of a carbon vacancy is missing, we restricted ourselves to the $\Gamma(\w)$ defined in  Eq.\ \eqref{eq:hybridization}.

\section{Summary and conclusion}
\label{sec:conclusion}
We investigated an effective two-orbital quantum impurity model 
for a carbon vacancy in locally curved monolayer of graphene.
Combining the estimations for the local impurity
parameters \cite{Yazyev2007,Cazalilla2012,Nanda2012} 
with the experimental STM data
\cite{Mao2017} we identified three different regimes, namely the
weak, intermediate, and strong hybridization regime.  
Our NRG calculations are carried out for
$T=4.2\,\mathrm{K}$ matching the experimental temperature \cite{Mao2017}.
The intermediate regime is characterized by SC fixed
points for both n and p doping which crosses over into unstable LM regimes
for a constant temperature when $\mu$ approaches the Dirac point
as the vanishing DOS of graphene exponentially suppresses
screening of the impurity spins. The other two 
regimes only show a Kondo effect in the p doped region
for large hybridization (curvature) or in the n doped region
for small hybridization (curvature).
For positive $\mu$ and significantly strong hybridization 
the systems is driven into a FI regime whose striking
feature is a single peak in the spectral function shifting away from
$\omega=0$ with increasing $\mu$. 

All three regimes are connected by changing the
graphene curvature that also changes the effective single particle
orbital energies: Not only the $\pi$ conduction band hybridizes with the
$\sigma$ orbital but also the vacancy induced bound $\pi$ state.
This physical mechanism driving 
the transition from weak to intermediate and finally to the strong
hybridization regime is related to the local
rippling of graphene that has been verified experimentally \cite{Mao2017}.
All three regimes can be understood in terms of 
local spin and charge configurations of a two-orbital impurity model.

Our findings are in qualitative agreement with recent experimental
studies by Mao et al.\ \cite{Mao2017} where scanning tunneling microscope (STM)
measurements on irradiated graphene sheets found all three scenarios albeit
with higher Kondo temperatures. We have shown that all three regimes are 
an intrinsic feature of our
realistic two-orbital model and are not necessarily dependent on the
exact form of the DOS as long as the pseudo-gap slope for $\omega
\rightarrow 0$ is present. Even for particularly small Coulomb
interactions we can still reproduce the intermediate and strong
hybridization regime
although $T_K$ is diminished substantially.  

The only drawback of our approach is the relatively low Kondo temperature
compared to the experiment. In order to predict larger $T_K$s,  
a substantial local hybridization is required
\cite{Vojta2010,MitchellFritz2013} whose microscopic origin is unclear: Note that the nearest neighbor 
hopping matrix element $t$ in pristine graphene has been approximate $2.9{\rm eV}$,
and hybridization matrix elements of the same order are 
required to achieve $T_K$s of the order of $60-80\,\mathrm{K}$.
This might be caused by several aspects not included in our model.
Firstly, in all calculations we assume that the Dirac Fermion DOS
prevails in the vicinity of the vacancy up to the bound state.
However, Kondo temperatures depend sensitively on the local $\Gamma(\w)$ close to the Dirac point.
The true energy dependence of $\Gamma(\w)$ in a curved environment -- a result of the vacancy, the 
substrate, and the high gate voltages -- might differ slightly from that of pristine graphene.
Secondly, we assumed that the STS is directly proportional to the spectral function of the $\sigma$-orbital
while in reality, the STM tip can locally couple to several different orbitals \cite{SchillerHershfield2000a}.
The additional asymmetries of such a resulting spectrum stemming from
the interference between multiple transport channels, one to the local
$\sigma$ orbital and
the others to the graphene p band, might lead to a larger value of $T_K$ by a
two-parameter Fano fit.
Thirdly, local rippling of graphene is caused by an instability of the
ideal 2D graphene sheet. Local curvatures will modify the phonon
spectrum, and it is expected that there is a significant electron
phonon coupling contributing to the hybridization. This electron
phonon interaction has been proven to enhance the effective
hybridization \cite{Cornaglia2005} and therefore $T_K$. In addition,
one expects inelastic contributions to the tunnel current if such a
breathing mode leads to an oscillation of the distance between the
vacancy and the STM tip. Such an additional contribution could modify
the overall shape of STS and therefore also change the FWHM used to
estimate $T_K$. 

\begin{acknowledgments}
We thank R.\ Bulla and M.\ Vojta for fruitful discussions.
We acknowledge support from the 
Deutsche Forschungsgemeinschaft via project AN-275/8-1 (F.B.A. and D.M.),
DOE-FG02-99ER45742 (E.Y.A. and J.M.), NSF DMR 1708158 (Y.J.),
Ministry of Science and Technology and also Academia Sinica of Taiwan (G.Y.G. and P.W.L.).

\end{acknowledgments}


%


\begin{thebibliography}{58}%
\makeatletter
\providecommand \@ifxundefined [1]{%
 \@ifx{#1\undefined}
}%
\providecommand \@ifnum [1]{%
 \ifnum #1\expandafter \@firstoftwo
 \else \expandafter \@secondoftwo
 \fi
}%
\providecommand \@ifx [1]{%
 \ifx #1\expandafter \@firstoftwo
 \else \expandafter \@secondoftwo
 \fi
}%
\providecommand \natexlab [1]{#1}%
\providecommand \enquote  [1]{``#1''}%
\providecommand \bibnamefont  [1]{#1}%
\providecommand \bibfnamefont [1]{#1}%
\providecommand \citenamefont [1]{#1}%
\providecommand \href@noop [0]{\@secondoftwo}%
\providecommand \href [0]{\begingroup \@sanitize@url \@href}%
\providecommand \@href[1]{\@@startlink{#1}\@@href}%
\providecommand \@@href[1]{\endgroup#1\@@endlink}%
\providecommand \@sanitize@url [0]{\catcode `\\12\catcode `\$12\catcode
  `\&12\catcode `\#12\catcode `\^12\catcode `\_12\catcode `\%12\relax}%
\providecommand \@@startlink[1]{}%
\providecommand \@@endlink[0]{}%
\providecommand \url  [0]{\begingroup\@sanitize@url \@url }%
\providecommand \@url [1]{\endgroup\@href {#1}{\urlprefix }}%
\providecommand \urlprefix  [0]{URL }%
\providecommand \Eprint [0]{\href }%
\providecommand \doibase [0]{http://dx.doi.org/}%
\providecommand \selectlanguage [0]{\@gobble}%
\providecommand \bibinfo  [0]{\@secondoftwo}%
\providecommand \bibfield  [0]{\@secondoftwo}%
\providecommand \translation [1]{[#1]}%
\providecommand \BibitemOpen [0]{}%
\providecommand \bibitemStop [0]{}%
\providecommand \bibitemNoStop [0]{.\EOS\space}%
\providecommand \EOS [0]{\spacefactor3000\relax}%
\providecommand \BibitemShut  [1]{\csname bibitem#1\endcsname}%
\let\auto@bib@innerbib\@empty
\bibitem [{\citenamefont {Novoselov}\ \emph {et~al.}(2005)\citenamefont
  {Novoselov}, \citenamefont {Geim}, \citenamefont {Morozov}, \citenamefont
  {Jiang}, \citenamefont {Katsnelson}, \citenamefont {Grigorieva},
  \citenamefont {Dubonos},\ and\ \citenamefont {Firsov}}]{Novoselov2005}%
  \BibitemOpen
  \bibfield  {author} {\bibinfo {author} {\bibfnamefont {K.~S.}\ \bibnamefont
  {Novoselov}}, \bibinfo {author} {\bibfnamefont {A.~K.}\ \bibnamefont {Geim}},
  \bibinfo {author} {\bibfnamefont {S.~V.}\ \bibnamefont {Morozov}}, \bibinfo
  {author} {\bibfnamefont {D.}~\bibnamefont {Jiang}}, \bibinfo {author}
  {\bibfnamefont {M.~I.}\ \bibnamefont {Katsnelson}}, \bibinfo {author}
  {\bibfnamefont {I.~V.}\ \bibnamefont {Grigorieva}}, \bibinfo {author}
  {\bibfnamefont {S.~V.}\ \bibnamefont {Dubonos}}, \ and\ \bibinfo {author}
  {\bibfnamefont {A.~A.}\ \bibnamefont {Firsov}},\ }\href {\doibase
  10.1038/nature04233} {\bibfield  {journal} {\bibinfo  {journal} {Nature}\
  }\textbf {\bibinfo {volume} {438}},\ \bibinfo {pages} {197} (\bibinfo {year}
  {2005})}\BibitemShut {NoStop}%
\bibitem [{\citenamefont {Castro~Neto}\ \emph {et~al.}(2009)\citenamefont
  {Castro~Neto}, \citenamefont {Guinea}, \citenamefont {Peres}, \citenamefont
  {Novoselov},\ and\ \citenamefont {Geim}}]{Neto2009}%
  \BibitemOpen
  \bibfield  {author} {\bibinfo {author} {\bibfnamefont {A.~H.}\ \bibnamefont
  {Castro~Neto}}, \bibinfo {author} {\bibfnamefont {F.}~\bibnamefont {Guinea}},
  \bibinfo {author} {\bibfnamefont {N.~M.~R.}\ \bibnamefont {Peres}}, \bibinfo
  {author} {\bibfnamefont {K.~S.}\ \bibnamefont {Novoselov}}, \ and\ \bibinfo
  {author} {\bibfnamefont {A.~K.}\ \bibnamefont {Geim}},\ }\href {\doibase
  10.1103/RevModPhys.81.109} {\bibfield  {journal} {\bibinfo  {journal} {Rev.
  Mod. Phys.}\ }\textbf {\bibinfo {volume} {81}},\ \bibinfo {pages} {109}
  (\bibinfo {year} {2009})}\BibitemShut {NoStop}%
\bibitem [{\citenamefont {Wallace}(1947)}]{Wallace1947}%
  \BibitemOpen
  \bibfield  {author} {\bibinfo {author} {\bibfnamefont {P.~R.}\ \bibnamefont
  {Wallace}},\ }\href {\doibase 10.1103/PhysRev.71.622} {\bibfield  {journal}
  {\bibinfo  {journal} {Phys. Rev.}\ }\textbf {\bibinfo {volume} {71}},\
  \bibinfo {pages} {622} (\bibinfo {year} {1947})}\BibitemShut {NoStop}%
\bibitem [{\citenamefont {Wang}\ \emph {et~al.}(2009)\citenamefont {Wang},
  \citenamefont {Huang}, \citenamefont {Song}, \citenamefont {Zhang},
  \citenamefont {Ma}, \citenamefont {Liang},\ and\ \citenamefont
  {Chen}}]{Wang2009}%
  \BibitemOpen
  \bibfield  {author} {\bibinfo {author} {\bibfnamefont {Y.}~\bibnamefont
  {Wang}}, \bibinfo {author} {\bibfnamefont {Y.}~\bibnamefont {Huang}},
  \bibinfo {author} {\bibfnamefont {Y.}~\bibnamefont {Song}}, \bibinfo {author}
  {\bibfnamefont {X.}~\bibnamefont {Zhang}}, \bibinfo {author} {\bibfnamefont
  {Y.}~\bibnamefont {Ma}}, \bibinfo {author} {\bibfnamefont {J.}~\bibnamefont
  {Liang}}, \ and\ \bibinfo {author} {\bibfnamefont {Y.}~\bibnamefont {Chen}},\
  }\href {\doibase 10.1021/nl802810g} {\bibfield  {journal} {\bibinfo
  {journal} {Nano Letters}\ }\textbf {\bibinfo {volume} {9}},\ \bibinfo {pages}
  {220} (\bibinfo {year} {2009})}\BibitemShut {NoStop}%
\bibitem [{\citenamefont {Nair}\ \emph {et~al.}(2012)\citenamefont {Nair},
  \citenamefont {Sepioni}, \citenamefont {Tsai}, \citenamefont {Lehtinen},
  \citenamefont {Keinonen}, \citenamefont {Krasheninnikov}, \citenamefont
  {Thomson}, \citenamefont {Geim},\ and\ \citenamefont
  {Grigorieva}}]{Nair2012}%
  \BibitemOpen
  \bibfield  {author} {\bibinfo {author} {\bibfnamefont {R.~R.}\ \bibnamefont
  {Nair}}, \bibinfo {author} {\bibfnamefont {M.}~\bibnamefont {Sepioni}},
  \bibinfo {author} {\bibfnamefont {I.-L.}\ \bibnamefont {Tsai}}, \bibinfo
  {author} {\bibfnamefont {O.}~\bibnamefont {Lehtinen}}, \bibinfo {author}
  {\bibfnamefont {J.}~\bibnamefont {Keinonen}}, \bibinfo {author}
  {\bibfnamefont {A.~V.}\ \bibnamefont {Krasheninnikov}}, \bibinfo {author}
  {\bibfnamefont {T.}~\bibnamefont {Thomson}}, \bibinfo {author} {\bibfnamefont
  {A.~K.}\ \bibnamefont {Geim}}, \ and\ \bibinfo {author} {\bibfnamefont
  {I.~V.}\ \bibnamefont {Grigorieva}},\ }\href {\doibase 10.1038/nphys2183}
  {\bibfield  {journal} {\bibinfo  {journal} {Nat Phys}\ }\textbf {\bibinfo
  {volume} {8}},\ \bibinfo {pages} {199} (\bibinfo {year} {2012})}\BibitemShut
  {NoStop}%
\bibitem [{\citenamefont {Nair}\ \emph {et~al.}(2013)\citenamefont {Nair},
  \citenamefont {Tsai}, \citenamefont {Sepioni}, \citenamefont {Lehtinen},
  \citenamefont {Keinonen}, \citenamefont {Krasheninnikov}, \citenamefont
  {Castro~Neto}, \citenamefont {Katsnelson}, \citenamefont {Geim},\ and\
  \citenamefont {Grigorieva}}]{Nair2013}%
  \BibitemOpen
  \bibfield  {author} {\bibinfo {author} {\bibfnamefont {R.~R.}\ \bibnamefont
  {Nair}}, \bibinfo {author} {\bibfnamefont {I.-L.}\ \bibnamefont {Tsai}},
  \bibinfo {author} {\bibfnamefont {M.}~\bibnamefont {Sepioni}}, \bibinfo
  {author} {\bibfnamefont {O.}~\bibnamefont {Lehtinen}}, \bibinfo {author}
  {\bibfnamefont {J.}~\bibnamefont {Keinonen}}, \bibinfo {author}
  {\bibfnamefont {A.~V.}\ \bibnamefont {Krasheninnikov}}, \bibinfo {author}
  {\bibfnamefont {A.~H.}\ \bibnamefont {Castro~Neto}}, \bibinfo {author}
  {\bibfnamefont {M.~I.}\ \bibnamefont {Katsnelson}}, \bibinfo {author}
  {\bibfnamefont {A.~K.}\ \bibnamefont {Geim}}, \ and\ \bibinfo {author}
  {\bibfnamefont {I.~V.}\ \bibnamefont {Grigorieva}},\ }\href
  {http://dx.doi.org/10.1038/ncomms3010} {\bibfield  {journal} {\bibinfo
  {journal} {Nature Communications}\ }\textbf {\bibinfo {volume} {4}},\
  \bibinfo {pages} {2010 EP } (\bibinfo {year} {2013})},\ \bibinfo {note}
  {article}\BibitemShut {NoStop}%
\bibitem [{\citenamefont {Chen}\ \emph {et~al.}(2011)\citenamefont {Chen},
  \citenamefont {Li}, \citenamefont {Cullen}, \citenamefont {Williams},\ and\
  \citenamefont {Fuhrer}}]{Chen2011}%
  \BibitemOpen
  \bibfield  {author} {\bibinfo {author} {\bibfnamefont {J.-H.}\ \bibnamefont
  {Chen}}, \bibinfo {author} {\bibfnamefont {L.}~\bibnamefont {Li}}, \bibinfo
  {author} {\bibfnamefont {W.~G.}\ \bibnamefont {Cullen}}, \bibinfo {author}
  {\bibfnamefont {E.~D.}\ \bibnamefont {Williams}}, \ and\ \bibinfo {author}
  {\bibfnamefont {M.~S.}\ \bibnamefont {Fuhrer}},\ }\href {\doibase
  10.1038/nphys1962} {\bibfield  {journal} {\bibinfo  {journal} {Nat Phys}\
  }\textbf {\bibinfo {volume} {7}},\ \bibinfo {pages} {535} (\bibinfo {year}
  {2011})}\BibitemShut {NoStop}%
\bibitem [{\citenamefont {Kane}\ and\ \citenamefont {Mele}(2005)}]{Kane2005}%
  \BibitemOpen
  \bibfield  {author} {\bibinfo {author} {\bibfnamefont {C.~L.}\ \bibnamefont
  {Kane}}\ and\ \bibinfo {author} {\bibfnamefont {E.~J.}\ \bibnamefont
  {Mele}},\ }\href {\doibase 10.1103/PhysRevLett.95.226801} {\bibfield
  {journal} {\bibinfo  {journal} {Phys. Rev. Lett.}\ }\textbf {\bibinfo
  {volume} {95}},\ \bibinfo {pages} {226801} (\bibinfo {year}
  {2005})}\BibitemShut {NoStop}%
\bibitem [{\citenamefont {Cho}\ \emph {et~al.}(2007)\citenamefont {Cho},
  \citenamefont {Chen},\ and\ \citenamefont {Fuhrer}}]{Cho2007}%
  \BibitemOpen
  \bibfield  {author} {\bibinfo {author} {\bibfnamefont {S.}~\bibnamefont
  {Cho}}, \bibinfo {author} {\bibfnamefont {Y.-F.}\ \bibnamefont {Chen}}, \
  and\ \bibinfo {author} {\bibfnamefont {M.~S.}\ \bibnamefont {Fuhrer}},\
  }\href {\doibase 10.1063/1.2784934} {\bibfield  {journal} {\bibinfo
  {journal} {Applied Physics Letters}\ }\textbf {\bibinfo {volume} {91}},\
  \bibinfo {pages} {123105} (\bibinfo {year} {2007})}\BibitemShut {NoStop}%
\bibitem [{\citenamefont {Vojta}\ \emph {et~al.}(2010)\citenamefont {Vojta},
  \citenamefont {Fritz},\ and\ \citenamefont {Bulla}}]{Vojta2010}%
  \BibitemOpen
  \bibfield  {author} {\bibinfo {author} {\bibfnamefont {M.}~\bibnamefont
  {Vojta}}, \bibinfo {author} {\bibfnamefont {L.}~\bibnamefont {Fritz}}, \ and\
  \bibinfo {author} {\bibfnamefont {R.}~\bibnamefont {Bulla}},\ }\href
  {http://stacks.iop.org/0295-5075/90/i=2/a=27006} {\bibfield  {journal}
  {\bibinfo  {journal} {EPL (Europhysics Letters)}\ }\textbf {\bibinfo {volume}
  {90}},\ \bibinfo {pages} {27006} (\bibinfo {year} {2010})}\BibitemShut
  {NoStop}%
\bibitem [{\citenamefont {Fritz}\ and\ \citenamefont
  {Vojta}(2013)}]{Fritz2013}%
  \BibitemOpen
  \bibfield  {author} {\bibinfo {author} {\bibfnamefont {L.}~\bibnamefont
  {Fritz}}\ and\ \bibinfo {author} {\bibfnamefont {M.}~\bibnamefont {Vojta}},\
  }\href {http://stacks.iop.org/0034-4885/76/i=3/a=032501} {\bibfield
  {journal} {\bibinfo  {journal} {Reports on Progress in Physics}\ }\textbf
  {\bibinfo {volume} {76}},\ \bibinfo {pages} {032501} (\bibinfo {year}
  {2013})}\BibitemShut {NoStop}%
\bibitem [{\citenamefont {Withoff}\ and\ \citenamefont
  {Fradkin}(1990)}]{WithoffFradkin1990}%
  \BibitemOpen
  \bibfield  {author} {\bibinfo {author} {\bibfnamefont {D.}~\bibnamefont
  {Withoff}}\ and\ \bibinfo {author} {\bibfnamefont {E.}~\bibnamefont
  {Fradkin}},\ }\href {\doibase 10.1103/PhysRevLett.64.1835} {\bibfield
  {journal} {\bibinfo  {journal} {Phys. Rev. Lett.}\ }\textbf {\bibinfo
  {volume} {64}},\ \bibinfo {pages} {1835} (\bibinfo {year}
  {1990})}\BibitemShut {NoStop}%
\bibitem [{\citenamefont {Gonzalez-Buxton}\ and\ \citenamefont
  {Ingersent}(1998)}]{GonzalezBuxtonIngersent1998}%
  \BibitemOpen
  \bibfield  {author} {\bibinfo {author} {\bibfnamefont {C.}~\bibnamefont
  {Gonzalez-Buxton}}\ and\ \bibinfo {author} {\bibfnamefont {K.}~\bibnamefont
  {Ingersent}},\ }\href {\doibase 10.1103/PhysRevB.57.14254} {\bibfield
  {journal} {\bibinfo  {journal} {Phys. Rev. B}\ }\textbf {\bibinfo {volume}
  {57}},\ \bibinfo {pages} {14254} (\bibinfo {year} {1998})}\BibitemShut
  {NoStop}%
\bibitem [{\citenamefont {Vojta}(2006)}]{Vojta2006}%
  \BibitemOpen
  \bibfield  {author} {\bibinfo {author} {\bibfnamefont {M.}~\bibnamefont
  {Vojta}},\ }\href {\doibase 10.1080/14786430500070396} {\bibfield  {journal}
  {\bibinfo  {journal} {Philosophical Magazine}\ }\textbf {\bibinfo {volume}
  {86}},\ \bibinfo {pages} {1807} (\bibinfo {year} {2006})},\ \Eprint
  {http://arxiv.org/abs/http://dx.doi.org/10.1080/14786430500070396}
  {http://dx.doi.org/10.1080/14786430500070396} \BibitemShut {NoStop}%
\bibitem [{\citenamefont {Rudenko}\ \emph {et~al.}(2012)\citenamefont
  {Rudenko}, \citenamefont {Keil}, \citenamefont {Katsnelson},\ and\
  \citenamefont {Lichtenstein}}]{CobaltOnGrapheneLDA2012}%
  \BibitemOpen
  \bibfield  {author} {\bibinfo {author} {\bibfnamefont {A.~N.}\ \bibnamefont
  {Rudenko}}, \bibinfo {author} {\bibfnamefont {F.~J.}\ \bibnamefont {Keil}},
  \bibinfo {author} {\bibfnamefont {M.~I.}\ \bibnamefont {Katsnelson}}, \ and\
  \bibinfo {author} {\bibfnamefont {A.~I.}\ \bibnamefont {Lichtenstein}},\
  }\href {\doibase 10.1103/PhysRevB.86.075422} {\bibfield  {journal} {\bibinfo
  {journal} {Phys. Rev. B}\ }\textbf {\bibinfo {volume} {86}},\ \bibinfo
  {pages} {075422} (\bibinfo {year} {2012})}\BibitemShut {NoStop}%
\bibitem [{\citenamefont {Mitchell}\ and\ \citenamefont
  {Fritz}(2013)}]{MitchellFritz2013}%
  \BibitemOpen
  \bibfield  {author} {\bibinfo {author} {\bibfnamefont {A.~K.}\ \bibnamefont
  {Mitchell}}\ and\ \bibinfo {author} {\bibfnamefont {L.}~\bibnamefont
  {Fritz}},\ }\href {\doibase 10.1103/PhysRevB.88.075104} {\bibfield  {journal}
  {\bibinfo  {journal} {Phys. Rev. B}\ }\textbf {\bibinfo {volume} {88}},\
  \bibinfo {pages} {075104} (\bibinfo {year} {2013})}\BibitemShut {NoStop}%
\bibitem [{\citenamefont {Ren}\ \emph {et~al.}(2014)\citenamefont {Ren},
  \citenamefont {Guo}, \citenamefont {Pan}, \citenamefont {Zhang},
  \citenamefont {Wu}, \citenamefont {Luo}, \citenamefont {Du}, \citenamefont
  {Pantelides},\ and\ \citenamefont {Gao}}]{RenCoGraphene2014}%
  \BibitemOpen
  \bibfield  {author} {\bibinfo {author} {\bibfnamefont {J.}~\bibnamefont
  {Ren}}, \bibinfo {author} {\bibfnamefont {H.}~\bibnamefont {Guo}}, \bibinfo
  {author} {\bibfnamefont {J.}~\bibnamefont {Pan}}, \bibinfo {author}
  {\bibfnamefont {Y.~Y.}\ \bibnamefont {Zhang}}, \bibinfo {author}
  {\bibfnamefont {X.}~\bibnamefont {Wu}}, \bibinfo {author} {\bibfnamefont
  {H.-G.}\ \bibnamefont {Luo}}, \bibinfo {author} {\bibfnamefont
  {S.}~\bibnamefont {Du}}, \bibinfo {author} {\bibfnamefont {S.~T.}\
  \bibnamefont {Pantelides}}, \ and\ \bibinfo {author} {\bibfnamefont {H.-J.}\
  \bibnamefont {Gao}},\ }\href {\doibase 10.1021/nl501425n} {\bibfield
  {journal} {\bibinfo  {journal} {Nano Letters}\ }\textbf {\bibinfo {volume}
  {14}},\ \bibinfo {pages} {4011} (\bibinfo {year} {2014})},\ \bibinfo {note}
  {pMID: 24905855},\ \Eprint
  {http://arxiv.org/abs/http://dx.doi.org/10.1021/nl501425n}
  {http://dx.doi.org/10.1021/nl501425n} \BibitemShut {NoStop}%
\bibitem [{\citenamefont {Pereira}\ \emph {et~al.}(2007)\citenamefont
  {Pereira}, \citenamefont {Nilsson},\ and\ \citenamefont
  {Castro~Neto}}]{Pereira2007}%
  \BibitemOpen
  \bibfield  {author} {\bibinfo {author} {\bibfnamefont {V.~M.}\ \bibnamefont
  {Pereira}}, \bibinfo {author} {\bibfnamefont {J.}~\bibnamefont {Nilsson}}, \
  and\ \bibinfo {author} {\bibfnamefont {A.~H.}\ \bibnamefont {Castro~Neto}},\
  }\href {\doibase 10.1103/PhysRevLett.99.166802} {\bibfield  {journal}
  {\bibinfo  {journal} {Phys. Rev. Lett.}\ }\textbf {\bibinfo {volume} {99}},\
  \bibinfo {pages} {166802} (\bibinfo {year} {2007})}\BibitemShut {NoStop}%
\bibitem [{\citenamefont {Nanda}\ \emph {et~al.}(2012)\citenamefont {Nanda},
  \citenamefont {Sherafati}, \citenamefont {Popović},\ and\ \citenamefont
  {Satpathy}}]{Nanda2012}%
  \BibitemOpen
  \bibfield  {author} {\bibinfo {author} {\bibfnamefont {B.~R.~K.}\
  \bibnamefont {Nanda}}, \bibinfo {author} {\bibfnamefont {M.}~\bibnamefont
  {Sherafati}}, \bibinfo {author} {\bibfnamefont {Z.~S.}\ \bibnamefont
  {Popović}}, \ and\ \bibinfo {author} {\bibfnamefont {S.}~\bibnamefont
  {Satpathy}},\ }\href {http://stacks.iop.org/1367-2630/14/i=8/a=083004}
  {\bibfield  {journal} {\bibinfo  {journal} {New Journal of Physics}\ }\textbf
  {\bibinfo {volume} {14}},\ \bibinfo {pages} {083004} (\bibinfo {year}
  {2012})}\BibitemShut {NoStop}%
\bibitem [{\citenamefont {{Cazalilla}}\ \emph {et~al.}(2012)\citenamefont
  {{Cazalilla}}, \citenamefont {{Iucci}}, \citenamefont {{Guinea}},\ and\
  \citenamefont {{Castro Neto}}}]{Cazalilla2012}%
  \BibitemOpen
  \bibfield  {author} {\bibinfo {author} {\bibfnamefont {M.~A.}\ \bibnamefont
  {{Cazalilla}}}, \bibinfo {author} {\bibfnamefont {A.}~\bibnamefont
  {{Iucci}}}, \bibinfo {author} {\bibfnamefont {F.}~\bibnamefont {{Guinea}}}, \
  and\ \bibinfo {author} {\bibfnamefont {A.~H.}\ \bibnamefont {{Castro
  Neto}}},\ }\href@noop {} {\bibfield  {journal} {\bibinfo  {journal} {ArXiv
  e-prints}\ } (\bibinfo {year} {2012})},\ \Eprint
  {http://arxiv.org/abs/1207.3135} {arXiv:1207.3135 [cond-mat.str-el]}
  \BibitemShut {NoStop}%
\bibitem [{\citenamefont {Wilson}(1975)}]{Wilson1975}%
  \BibitemOpen
  \bibfield  {author} {\bibinfo {author} {\bibfnamefont {K.~G.}\ \bibnamefont
  {Wilson}},\ }\href {\doibase 10.1103/RevModPhys.47.773} {\bibfield  {journal}
  {\bibinfo  {journal} {Rev. Mod. Phys.}\ }\textbf {\bibinfo {volume} {47}},\
  \bibinfo {pages} {773} (\bibinfo {year} {1975})}\BibitemShut {NoStop}%
\bibitem [{\citenamefont {Bulla}\ \emph {et~al.}(2008)\citenamefont {Bulla},
  \citenamefont {Costi},\ and\ \citenamefont {Pruschke}}]{Bulla2008}%
  \BibitemOpen
  \bibfield  {author} {\bibinfo {author} {\bibfnamefont {R.}~\bibnamefont
  {Bulla}}, \bibinfo {author} {\bibfnamefont {T.~A.}\ \bibnamefont {Costi}}, \
  and\ \bibinfo {author} {\bibfnamefont {T.}~\bibnamefont {Pruschke}},\ }\href
  {\doibase 10.1103/RevModPhys.80.395} {\bibfield  {journal} {\bibinfo
  {journal} {Rev. Mod. Phys.}\ }\textbf {\bibinfo {volume} {80}},\ \bibinfo
  {pages} {395} (\bibinfo {year} {2008})}\BibitemShut {NoStop}%
\bibitem [{\citenamefont {Peres}\ \emph {et~al.}(2009)\citenamefont {Peres},
  \citenamefont {Tsai}, \citenamefont {Santos},\ and\ \citenamefont
  {Ribeiro}}]{Peres2009}%
  \BibitemOpen
  \bibfield  {author} {\bibinfo {author} {\bibfnamefont {N.~M.~R.}\
  \bibnamefont {Peres}}, \bibinfo {author} {\bibfnamefont {S.-W.}\ \bibnamefont
  {Tsai}}, \bibinfo {author} {\bibfnamefont {J.~E.}\ \bibnamefont {Santos}}, \
  and\ \bibinfo {author} {\bibfnamefont {R.~M.}\ \bibnamefont {Ribeiro}},\
  }\href {\doibase 10.1103/PhysRevB.79.155442} {\bibfield  {journal} {\bibinfo
  {journal} {Phys. Rev. B}\ }\textbf {\bibinfo {volume} {79}},\ \bibinfo
  {pages} {155442} (\bibinfo {year} {2009})}\BibitemShut {NoStop}%
\bibitem [{\citenamefont {Mao}\ \emph {et~al.}(2017)\citenamefont {Mao},
  \citenamefont {Jiang}, \citenamefont {Lo}, \citenamefont {May}, \citenamefont
  {Li}, \citenamefont {Guo}, \citenamefont {Anders}, \citenamefont {Taniguchi},
  \citenamefont {Watanabe},\ and\ \citenamefont {Andrei}}]{Mao2017}%
  \BibitemOpen
  \bibfield  {author} {\bibinfo {author} {\bibfnamefont {J.}~\bibnamefont
  {Mao}}, \bibinfo {author} {\bibfnamefont {Y.}~\bibnamefont {Jiang}}, \bibinfo
  {author} {\bibfnamefont {P.-W.}\ \bibnamefont {Lo}}, \bibinfo {author}
  {\bibfnamefont {D.}~\bibnamefont {May}}, \bibinfo {author} {\bibfnamefont
  {G.}~\bibnamefont {Li}}, \bibinfo {author} {\bibfnamefont {G.-Y.}\
  \bibnamefont {Guo}}, \bibinfo {author} {\bibfnamefont {F.}~\bibnamefont
  {Anders}}, \bibinfo {author} {\bibfnamefont {T.}~\bibnamefont {Taniguchi}},
  \bibinfo {author} {\bibfnamefont {K.}~\bibnamefont {Watanabe}}, \ and\
  \bibinfo {author} {\bibfnamefont {E.~Y.}\ \bibnamefont {Andrei}},\
  }\href@noop {} {\bibfield  {journal} {\bibinfo  {journal} {arXiv:1711.06942}\
  } (\bibinfo {year} {2017})}\BibitemShut {NoStop}%
\bibitem [{\citenamefont {Miranda}\ \emph {et~al.}(2014)\citenamefont
  {Miranda}, \citenamefont {Dias~da Silva},\ and\ \citenamefont
  {Lewenkopf}}]{Miranda2014}%
  \BibitemOpen
  \bibfield  {author} {\bibinfo {author} {\bibfnamefont {V.~G.}\ \bibnamefont
  {Miranda}}, \bibinfo {author} {\bibfnamefont {L.~G. G.~V.}\ \bibnamefont
  {Dias~da Silva}}, \ and\ \bibinfo {author} {\bibfnamefont {C.~H.}\
  \bibnamefont {Lewenkopf}},\ }\href {\doibase 10.1103/PhysRevB.90.201101}
  {\bibfield  {journal} {\bibinfo  {journal} {Phys. Rev. B}\ }\textbf {\bibinfo
  {volume} {90}},\ \bibinfo {pages} {201101} (\bibinfo {year}
  {2014})}\BibitemShut {NoStop}%
\bibitem [{\citenamefont {Ruiz-Tijerina}\ and\ \citenamefont {Dias~da
  Silva}(2017)}]{Tijerina2017}%
  \BibitemOpen
  \bibfield  {author} {\bibinfo {author} {\bibfnamefont {D.~A.}\ \bibnamefont
  {Ruiz-Tijerina}}\ and\ \bibinfo {author} {\bibfnamefont {L.~G. G.~V.}\
  \bibnamefont {Dias~da Silva}},\ }\href {\doibase 10.1103/PhysRevB.95.115408}
  {\bibfield  {journal} {\bibinfo  {journal} {Phys. Rev. B}\ }\textbf {\bibinfo
  {volume} {95}},\ \bibinfo {pages} {115408} (\bibinfo {year}
  {2017})}\BibitemShut {NoStop}%
\bibitem [{\citenamefont {Pereira}\ \emph {et~al.}(2008)\citenamefont
  {Pereira}, \citenamefont {Lopes~dos Santos},\ and\ \citenamefont
  {Castro~Neto}}]{Pereira2008}%
  \BibitemOpen
  \bibfield  {author} {\bibinfo {author} {\bibfnamefont {V.~M.}\ \bibnamefont
  {Pereira}}, \bibinfo {author} {\bibfnamefont {J.~M.~B.}\ \bibnamefont
  {Lopes~dos Santos}}, \ and\ \bibinfo {author} {\bibfnamefont {A.~H.}\
  \bibnamefont {Castro~Neto}},\ }\href {\doibase 10.1103/PhysRevB.77.115109}
  {\bibfield  {journal} {\bibinfo  {journal} {Phys. Rev. B}\ }\textbf {\bibinfo
  {volume} {77}},\ \bibinfo {pages} {115109} (\bibinfo {year}
  {2008})}\BibitemShut {NoStop}%
\bibitem [{\citenamefont {Ole\ifmmode~\acute{s}\else
  \'{s}\fi{}}(1983)}]{Oles1983}%
  \BibitemOpen
  \bibfield  {author} {\bibinfo {author} {\bibfnamefont {A.~M.}\ \bibnamefont
  {Ole\ifmmode~\acute{s}\else \'{s}\fi{}}},\ }\href {\doibase
  10.1103/PhysRevB.28.327} {\bibfield  {journal} {\bibinfo  {journal} {Phys.
  Rev. B}\ }\textbf {\bibinfo {volume} {28}},\ \bibinfo {pages} {327} (\bibinfo
  {year} {1983})}\BibitemShut {NoStop}%
\bibitem [{\citenamefont {{Reich, S. and Maultzsch, J. and Thomsen, C. and
  Ordej\'on, P.}}(2002)}]{Reich2002}%
  \BibitemOpen
  \bibfield  {author} {\bibinfo {author} {\bibnamefont {{Reich, S. and
  Maultzsch, J. and Thomsen, C. and Ordej\'on, P.}}},\ }\href {\doibase
  10.1103/PhysRevB.66.035412} {\bibfield  {journal} {\bibinfo  {journal} {Phys.
  Rev. B}\ }\textbf {\bibinfo {volume} {66}},\ \bibinfo {pages} {035412}
  (\bibinfo {year} {2002})}\BibitemShut {NoStop}%
\bibitem [{\citenamefont {Nanda}\ and\ \citenamefont
  {Satpathy}(2009)}]{Nanda2009}%
  \BibitemOpen
  \bibfield  {author} {\bibinfo {author} {\bibfnamefont {B.~R.~K.}\
  \bibnamefont {Nanda}}\ and\ \bibinfo {author} {\bibfnamefont
  {S.}~\bibnamefont {Satpathy}},\ }\href {\doibase 10.1103/PhysRevB.80.165430}
  {\bibfield  {journal} {\bibinfo  {journal} {Phys. Rev. B}\ }\textbf {\bibinfo
  {volume} {80}},\ \bibinfo {pages} {165430} (\bibinfo {year}
  {2009})}\BibitemShut {NoStop}%
\bibitem [{\citenamefont {Yazyev}\ and\ \citenamefont
  {Helm}(2007)}]{Yazyev2007}%
  \BibitemOpen
  \bibfield  {author} {\bibinfo {author} {\bibfnamefont {O.~V.}\ \bibnamefont
  {Yazyev}}\ and\ \bibinfo {author} {\bibfnamefont {L.}~\bibnamefont {Helm}},\
  }\href {\doibase 10.1103/PhysRevB.75.125408} {\bibfield  {journal} {\bibinfo
  {journal} {Phys. Rev. B}\ }\textbf {\bibinfo {volume} {75}},\ \bibinfo
  {pages} {125408} (\bibinfo {year} {2007})}\BibitemShut {NoStop}%
\bibitem [{\citenamefont {Padmanabhan}\ and\ \citenamefont
  {Nanda}(2016)}]{Padmanabhan2016}%
  \BibitemOpen
  \bibfield  {author} {\bibinfo {author} {\bibfnamefont {H.}~\bibnamefont
  {Padmanabhan}}\ and\ \bibinfo {author} {\bibfnamefont {B.~R.~K.}\
  \bibnamefont {Nanda}},\ }\href {\doibase 10.1103/PhysRevB.93.165403}
  {\bibfield  {journal} {\bibinfo  {journal} {Phys. Rev. B}\ }\textbf {\bibinfo
  {volume} {93}},\ \bibinfo {pages} {165403} (\bibinfo {year}
  {2016})}\BibitemShut {NoStop}%
\bibitem [{\citenamefont {Miranda}\ \emph {et~al.}(2016)\citenamefont
  {Miranda}, \citenamefont {Dias~da Silva},\ and\ \citenamefont
  {Lewenkopf}}]{UestimateMiranda2016}%
  \BibitemOpen
  \bibfield  {author} {\bibinfo {author} {\bibfnamefont {V.~G.}\ \bibnamefont
  {Miranda}}, \bibinfo {author} {\bibfnamefont {L.~G. G.~V.}\ \bibnamefont
  {Dias~da Silva}}, \ and\ \bibinfo {author} {\bibfnamefont {C.~H.}\
  \bibnamefont {Lewenkopf}},\ }\href {\doibase 10.1103/PhysRevB.94.075114}
  {\bibfield  {journal} {\bibinfo  {journal} {Phys. Rev. B}\ }\textbf {\bibinfo
  {volume} {94}},\ \bibinfo {pages} {075114} (\bibinfo {year}
  {2016})}\BibitemShut {NoStop}%
\bibitem [{\citenamefont {Kanao}\ \emph {et~al.}(2012)\citenamefont {Kanao},
  \citenamefont {Matsuura},\ and\ \citenamefont {Ogata}}]{Kanao2012}%
  \BibitemOpen
  \bibfield  {author} {\bibinfo {author} {\bibfnamefont {T.}~\bibnamefont
  {Kanao}}, \bibinfo {author} {\bibfnamefont {H.}~\bibnamefont {Matsuura}}, \
  and\ \bibinfo {author} {\bibfnamefont {M.}~\bibnamefont {Ogata}},\ }\href
  {\doibase 10.1143/JPSJ.81.063709} {\bibfield  {journal} {\bibinfo  {journal}
  {Journal of the Physical Society of Japan}\ }\textbf {\bibinfo {volume}
  {81}},\ \bibinfo {pages} {063709} (\bibinfo {year} {2012})}\BibitemShut
  {NoStop}%
\bibitem [{\citenamefont {Hewson}(1993)}]{Hewson1993}%
  \BibitemOpen
  \bibfield  {author} {\bibinfo {author} {\bibfnamefont {A.~C.}\ \bibnamefont
  {Hewson}},\ }\href {\doibase 10.1017/CBO9780511470752} {\emph {\bibinfo
  {title} {The Kondo Problem to Heavy Fermions}}},\ Cambridge Studies in
  Magnetism\ (\bibinfo  {publisher} {Cambridge University Press},\ \bibinfo
  {year} {1993})\BibitemShut {NoStop}%
\bibitem [{\citenamefont {Chen}\ and\ \citenamefont
  {Jayaprakash}(1995)}]{Chen1995}%
  \BibitemOpen
  \bibfield  {author} {\bibinfo {author} {\bibfnamefont {K.}~\bibnamefont
  {Chen}}\ and\ \bibinfo {author} {\bibfnamefont {C.}~\bibnamefont
  {Jayaprakash}},\ }\href {http://stacks.iop.org/0953-8984/7/i=37/a=003}
  {\bibfield  {journal} {\bibinfo  {journal} {Journal of Physics: Condensed
  Matter}\ }\textbf {\bibinfo {volume} {7}},\ \bibinfo {pages} {L491} (\bibinfo
  {year} {1995})}\BibitemShut {NoStop}%
\bibitem [{\citenamefont {Ingersent}(1996)}]{Ingersent1996}%
  \BibitemOpen
  \bibfield  {author} {\bibinfo {author} {\bibfnamefont {K.}~\bibnamefont
  {Ingersent}},\ }\href {\doibase 10.1103/PhysRevB.54.11936} {\bibfield
  {journal} {\bibinfo  {journal} {Phys. Rev. B}\ }\textbf {\bibinfo {volume}
  {54}},\ \bibinfo {pages} {11936} (\bibinfo {year} {1996})}\BibitemShut
  {NoStop}%
\bibitem [{\citenamefont {Bulla}\ \emph {et~al.}(1997)\citenamefont {Bulla},
  \citenamefont {Pruschke},\ and\ \citenamefont
  {Hewson}}]{BullaPruschkeHewson1997}%
  \BibitemOpen
  \bibfield  {author} {\bibinfo {author} {\bibfnamefont {R.}~\bibnamefont
  {Bulla}}, \bibinfo {author} {\bibfnamefont {T.}~\bibnamefont {Pruschke}}, \
  and\ \bibinfo {author} {\bibfnamefont {A.~C.}\ \bibnamefont {Hewson}},\
  }\href {http://stacks.iop.org/0953-8984/9/i=47/a=014} {\bibfield  {journal}
  {\bibinfo  {journal} {Journal of Physics: Condensed Matter}\ }\textbf
  {\bibinfo {volume} {9}},\ \bibinfo {pages} {10463} (\bibinfo {year}
  {1997})}\BibitemShut {NoStop}%
\bibitem [{\citenamefont {Krishna-murthy}\ \emph
  {et~al.}(1980{\natexlab{a}})\citenamefont {Krishna-murthy}, \citenamefont
  {Wilkins},\ and\ \citenamefont {Wilson}}]{Krishna-murthy1980a}%
  \BibitemOpen
  \bibfield  {author} {\bibinfo {author} {\bibfnamefont {H.~R.}\ \bibnamefont
  {Krishna-murthy}}, \bibinfo {author} {\bibfnamefont {J.~W.}\ \bibnamefont
  {Wilkins}}, \ and\ \bibinfo {author} {\bibfnamefont {K.~G.}\ \bibnamefont
  {Wilson}},\ }\href {\doibase 10.1103/PhysRevB.21.1003} {\bibfield  {journal}
  {\bibinfo  {journal} {Phys. Rev. B}\ }\textbf {\bibinfo {volume} {21}},\
  \bibinfo {pages} {1003} (\bibinfo {year} {1980}{\natexlab{a}})}\BibitemShut
  {NoStop}%
\bibitem [{\citenamefont {Krishna-murthy}\ \emph
  {et~al.}(1980{\natexlab{b}})\citenamefont {Krishna-murthy}, \citenamefont
  {Wilkins},\ and\ \citenamefont {Wilson}}]{Krishna-murthy1980b}%
  \BibitemOpen
  \bibfield  {author} {\bibinfo {author} {\bibfnamefont {H.~R.}\ \bibnamefont
  {Krishna-murthy}}, \bibinfo {author} {\bibfnamefont {J.~W.}\ \bibnamefont
  {Wilkins}}, \ and\ \bibinfo {author} {\bibfnamefont {K.~G.}\ \bibnamefont
  {Wilson}},\ }\href {\doibase 10.1103/PhysRevB.21.1044} {\bibfield  {journal}
  {\bibinfo  {journal} {Phys. Rev. B}\ }\textbf {\bibinfo {volume} {21}},\
  \bibinfo {pages} {1044} (\bibinfo {year} {1980}{\natexlab{b}})}\BibitemShut
  {NoStop}%
\bibitem [{\citenamefont {Anders}\ and\ \citenamefont
  {Schiller}(2005)}]{Anders2005}%
  \BibitemOpen
  \bibfield  {author} {\bibinfo {author} {\bibfnamefont {F.~B.}\ \bibnamefont
  {Anders}}\ and\ \bibinfo {author} {\bibfnamefont {A.}~\bibnamefont
  {Schiller}},\ }\href {\doibase 10.1103/PhysRevLett.95.196801} {\bibfield
  {journal} {\bibinfo  {journal} {Phys. Rev. Lett.}\ }\textbf {\bibinfo
  {volume} {95}},\ \bibinfo {pages} {196801} (\bibinfo {year}
  {2005})}\BibitemShut {NoStop}%
\bibitem [{\citenamefont {Anders}\ and\ \citenamefont
  {Schiller}(2006)}]{Anders2006}%
  \BibitemOpen
  \bibfield  {author} {\bibinfo {author} {\bibfnamefont {F.~B.}\ \bibnamefont
  {Anders}}\ and\ \bibinfo {author} {\bibfnamefont {A.}~\bibnamefont
  {Schiller}},\ }\href {\doibase 10.1103/PhysRevB.74.245113} {\bibfield
  {journal} {\bibinfo  {journal} {Phys. Rev. B}\ }\textbf {\bibinfo {volume}
  {74}},\ \bibinfo {pages} {245113} (\bibinfo {year} {2006})}\BibitemShut
  {NoStop}%
\bibitem [{\citenamefont {Weichselbaum}\ and\ \citenamefont {von
  Delft}(2007)}]{Weichselbaum2007}%
  \BibitemOpen
  \bibfield  {author} {\bibinfo {author} {\bibfnamefont {A.}~\bibnamefont
  {Weichselbaum}}\ and\ \bibinfo {author} {\bibfnamefont {J.}~\bibnamefont {von
  Delft}},\ }\href {\doibase 10.1103/PhysRevLett.99.076402} {\bibfield
  {journal} {\bibinfo  {journal} {Phys. Rev. Lett.}\ }\textbf {\bibinfo
  {volume} {99}},\ \bibinfo {pages} {076402} (\bibinfo {year}
  {2007})}\BibitemShut {NoStop}%
\bibitem [{\citenamefont {Peters}\ \emph {et~al.}(2006)\citenamefont {Peters},
  \citenamefont {Pruschke},\ and\ \citenamefont {Anders}}]{Peters2006}%
  \BibitemOpen
  \bibfield  {author} {\bibinfo {author} {\bibfnamefont {R.}~\bibnamefont
  {Peters}}, \bibinfo {author} {\bibfnamefont {T.}~\bibnamefont {Pruschke}}, \
  and\ \bibinfo {author} {\bibfnamefont {F.~B.}\ \bibnamefont {Anders}},\
  }\href {\doibase 10.1103/PhysRevB.74.245114} {\bibfield  {journal} {\bibinfo
  {journal} {Phys. Rev. B}\ }\textbf {\bibinfo {volume} {74}},\ \bibinfo
  {pages} {245114} (\bibinfo {year} {2006})}\BibitemShut {NoStop}%
\bibitem [{\citenamefont {Cox}\ and\ \citenamefont
  {Zawadowski}(1998)}]{CoxZawa98}%
  \BibitemOpen
  \bibfield  {author} {\bibinfo {author} {\bibfnamefont {D.~L.}\ \bibnamefont
  {Cox}}\ and\ \bibinfo {author} {\bibfnamefont {A.}~\bibnamefont
  {Zawadowski}},\ }\href@noop {} {\bibfield  {journal} {\bibinfo  {journal}
  {Advances in Physics}\ }\textbf {\bibinfo {volume} {47}},\ \bibinfo {pages}
  {599} (\bibinfo {year} {1998})},\ \bibinfo {note} {for a review on the
  multi-channel models}\BibitemShut {NoStop}%
\bibitem [{\citenamefont {Roch}\ \emph {et~al.}(2009)\citenamefont {Roch},
  \citenamefont {Florens}, \citenamefont {Costi}, \citenamefont {Wernsdorfer},\
  and\ \citenamefont {Balestro}}]{Roch2009}%
  \BibitemOpen
  \bibfield  {author} {\bibinfo {author} {\bibfnamefont {N.}~\bibnamefont
  {Roch}}, \bibinfo {author} {\bibfnamefont {S.}~\bibnamefont {Florens}},
  \bibinfo {author} {\bibfnamefont {T.~A.}\ \bibnamefont {Costi}}, \bibinfo
  {author} {\bibfnamefont {W.}~\bibnamefont {Wernsdorfer}}, \ and\ \bibinfo
  {author} {\bibfnamefont {F.}~\bibnamefont {Balestro}},\ }\href {\doibase
  10.1103/PhysRevLett.103.197202} {\bibfield  {journal} {\bibinfo  {journal}
  {Phys. Rev. Lett.}\ }\textbf {\bibinfo {volume} {103}},\ \bibinfo {pages}
  {197202} (\bibinfo {year} {2009})}\BibitemShut {NoStop}%
\bibitem [{\citenamefont {Goldhaber-Gordon}\ \emph {et~al.}(1998)\citenamefont
  {Goldhaber-Gordon}, \citenamefont {G\"ores}, \citenamefont {Kastner},
  \citenamefont {Shtrikman}, \citenamefont {Mahalu},\ and\ \citenamefont
  {Meirav}}]{GoldhaberGordon1998}%
  \BibitemOpen
  \bibfield  {author} {\bibinfo {author} {\bibfnamefont {D.}~\bibnamefont
  {Goldhaber-Gordon}}, \bibinfo {author} {\bibfnamefont {J.}~\bibnamefont
  {G\"ores}}, \bibinfo {author} {\bibfnamefont {M.~A.}\ \bibnamefont
  {Kastner}}, \bibinfo {author} {\bibfnamefont {H.}~\bibnamefont {Shtrikman}},
  \bibinfo {author} {\bibfnamefont {D.}~\bibnamefont {Mahalu}}, \ and\ \bibinfo
  {author} {\bibfnamefont {U.}~\bibnamefont {Meirav}},\ }\href {\doibase
  10.1103/PhysRevLett.81.5225} {\bibfield  {journal} {\bibinfo  {journal}
  {Phys. Rev. Lett.}\ }\textbf {\bibinfo {volume} {81}},\ \bibinfo {pages}
  {5225} (\bibinfo {year} {1998})}\BibitemShut {NoStop}%
\bibitem [{\citenamefont {Lo}\ \emph {et~al.}(2014)\citenamefont {Lo},
  \citenamefont {Guo},\ and\ \citenamefont {Anders}}]{Lo-graphene-2014}%
  \BibitemOpen
  \bibfield  {author} {\bibinfo {author} {\bibfnamefont {P.-W.}\ \bibnamefont
  {Lo}}, \bibinfo {author} {\bibfnamefont {G.-Y.}\ \bibnamefont {Guo}}, \ and\
  \bibinfo {author} {\bibfnamefont {F.~B.}\ \bibnamefont {Anders}},\ }\href
  {\doibase 10.1103/PhysRevB.89.195424} {\bibfield  {journal} {\bibinfo
  {journal} {Phys. Rev. B}\ }\textbf {\bibinfo {volume} {89}},\ \bibinfo
  {pages} {195424} (\bibinfo {year} {2014})}\BibitemShut {NoStop}%
\bibitem [{\citenamefont {Esat}\ \emph {et~al.}(2015)\citenamefont {Esat},
  \citenamefont {Deilmann}, \citenamefont {Lechtenberg}, \citenamefont
  {Wagner}, \citenamefont {Kr\"uger}, \citenamefont {Temirov}, \citenamefont
  {Anders}, \citenamefont {Rohlfing},\ and\ \citenamefont
  {Tautz}}]{AU-PTCDA-monomer}%
  \BibitemOpen
  \bibfield  {author} {\bibinfo {author} {\bibfnamefont {T.}~\bibnamefont
  {Esat}}, \bibinfo {author} {\bibfnamefont {T.}~\bibnamefont {Deilmann}},
  \bibinfo {author} {\bibfnamefont {B.}~\bibnamefont {Lechtenberg}}, \bibinfo
  {author} {\bibfnamefont {C.}~\bibnamefont {Wagner}}, \bibinfo {author}
  {\bibfnamefont {P.}~\bibnamefont {Kr\"uger}}, \bibinfo {author}
  {\bibfnamefont {R.}~\bibnamefont {Temirov}}, \bibinfo {author} {\bibfnamefont
  {F.~B.}\ \bibnamefont {Anders}}, \bibinfo {author} {\bibfnamefont
  {M.}~\bibnamefont {Rohlfing}}, \ and\ \bibinfo {author} {\bibfnamefont
  {F.~S.}\ \bibnamefont {Tautz}},\ }\href {\doibase 10.1103/PhysRevB.91.144415}
  {\bibfield  {journal} {\bibinfo  {journal} {Phys. Rev. B}\ }\textbf {\bibinfo
  {volume} {91}},\ \bibinfo {pages} {144415} (\bibinfo {year}
  {2015})}\BibitemShut {NoStop}%
\bibitem [{\citenamefont {Schiller}\ and\ \citenamefont
  {Hershfield}(2000)}]{SchillerHershfield2000a}%
  \BibitemOpen
  \bibfield  {author} {\bibinfo {author} {\bibfnamefont {A.}~\bibnamefont
  {Schiller}}\ and\ \bibinfo {author} {\bibfnamefont {S.}~\bibnamefont
  {Hershfield}},\ }\href@noop {} {\bibfield  {journal} {\bibinfo  {journal}
  {Phys. Rev. B}\ }\textbf {\bibinfo {volume} {61}},\ \bibinfo {pages} {9036}
  (\bibinfo {year} {2000})}\BibitemShut {NoStop}%
\bibitem [{\citenamefont {Wehling}\ \emph {et~al.}(2011)\citenamefont
  {Wehling}, \citenamefont {\ifmmode \mbox{\c{S}}\else \c{S}\fi{}a\ifmmode
  \mbox{\c{s}}\else \c{s}\fi{}\ifmmode \imath \else \i
  \fi{}o\ifmmode~\breve{g}\else \u{g}\fi{}lu}, \citenamefont {Friedrich},
  \citenamefont {Lichtenstein}, \citenamefont {Katsnelson},\ and\ \citenamefont
  {Bl\"ugel}}]{Wehling2011}%
  \BibitemOpen
  \bibfield  {author} {\bibinfo {author} {\bibfnamefont {T.~O.}\ \bibnamefont
  {Wehling}}, \bibinfo {author} {\bibfnamefont {E.}~\bibnamefont {\ifmmode
  \mbox{\c{S}}\else \c{S}\fi{}a\ifmmode \mbox{\c{s}}\else \c{s}\fi{}\ifmmode
  \imath \else \i \fi{}o\ifmmode~\breve{g}\else \u{g}\fi{}lu}}, \bibinfo
  {author} {\bibfnamefont {C.}~\bibnamefont {Friedrich}}, \bibinfo {author}
  {\bibfnamefont {A.~I.}\ \bibnamefont {Lichtenstein}}, \bibinfo {author}
  {\bibfnamefont {M.~I.}\ \bibnamefont {Katsnelson}}, \ and\ \bibinfo {author}
  {\bibfnamefont {S.}~\bibnamefont {Bl\"ugel}},\ }\href {\doibase
  10.1103/PhysRevLett.106.236805} {\bibfield  {journal} {\bibinfo  {journal}
  {Phys. Rev. Lett.}\ }\textbf {\bibinfo {volume} {106}},\ \bibinfo {pages}
  {236805} (\bibinfo {year} {2011})}\BibitemShut {NoStop}%
\bibitem [{\citenamefont {Verg\'es}\ \emph {et~al.}(2010)\citenamefont
  {Verg\'es}, \citenamefont {SanFabi\'an}, \citenamefont {Chiappe},\ and\
  \citenamefont {Louis}}]{Verges2010}%
  \BibitemOpen
  \bibfield  {author} {\bibinfo {author} {\bibfnamefont {J.~A.}\ \bibnamefont
  {Verg\'es}}, \bibinfo {author} {\bibfnamefont {E.}~\bibnamefont
  {SanFabi\'an}}, \bibinfo {author} {\bibfnamefont {G.}~\bibnamefont
  {Chiappe}}, \ and\ \bibinfo {author} {\bibfnamefont {E.}~\bibnamefont
  {Louis}},\ }\href {\doibase 10.1103/PhysRevB.81.085120} {\bibfield  {journal}
  {\bibinfo  {journal} {Phys. Rev. B}\ }\textbf {\bibinfo {volume} {81}},\
  \bibinfo {pages} {085120} (\bibinfo {year} {2010})}\BibitemShut {NoStop}%
\bibitem [{\citenamefont {Schrieffer}\ and\ \citenamefont
  {Wolff}(1966)}]{SchriefferWol66}%
  \BibitemOpen
  \bibfield  {author} {\bibinfo {author} {\bibfnamefont {J.~R.}\ \bibnamefont
  {Schrieffer}}\ and\ \bibinfo {author} {\bibfnamefont {P.~A.}\ \bibnamefont
  {Wolff}},\ }\href@noop {} {\bibfield  {journal} {\bibinfo  {journal} {Phys.
  Rev.}\ }\textbf {\bibinfo {volume} {149}},\ \bibinfo {pages} {491} (\bibinfo
  {year} {1966})}\BibitemShut {NoStop}%
\bibitem [{\citenamefont {Cornaglia}\ \emph {et~al.}(2005)\citenamefont
  {Cornaglia}, \citenamefont {Grempel},\ and\ \citenamefont
  {Ness}}]{Cornaglia2005}%
  \BibitemOpen
  \bibfield  {author} {\bibinfo {author} {\bibfnamefont {P.~S.}\ \bibnamefont
  {Cornaglia}}, \bibinfo {author} {\bibfnamefont {D.~R.}\ \bibnamefont
  {Grempel}}, \ and\ \bibinfo {author} {\bibfnamefont {H.}~\bibnamefont
  {Ness}},\ }\href {\doibase 10.1103/PhysRevB.71.075320} {\bibfield  {journal}
  {\bibinfo  {journal} {Phys. Rev. B}\ }\textbf {\bibinfo {volume} {71}},\
  \bibinfo {pages} {075320} (\bibinfo {year} {2005})}\BibitemShut {NoStop}%
\bibitem [{\citenamefont {Cornaglia}\ \emph {et~al.}(2011)\citenamefont
  {Cornaglia}, \citenamefont {Bas}, \citenamefont {Aligia},\ and\ \citenamefont
  {Balseiro}}]{Cornaglia2011}%
  \BibitemOpen
  \bibfield  {author} {\bibinfo {author} {\bibfnamefont {P.~S.}\ \bibnamefont
  {Cornaglia}}, \bibinfo {author} {\bibfnamefont {P.~R.}\ \bibnamefont {Bas}},
  \bibinfo {author} {\bibfnamefont {A.~A.}\ \bibnamefont {Aligia}}, \ and\
  \bibinfo {author} {\bibfnamefont {C.~A.}\ \bibnamefont {Balseiro}},\ }\href
  {http://stacks.iop.org/0295-5075/93/i=4/a=47005} {\bibfield  {journal}
  {\bibinfo  {journal} {EPL (Europhysics Letters)}\ }\textbf {\bibinfo {volume}
  {93}},\ \bibinfo {pages} {47005} (\bibinfo {year} {2011})}\BibitemShut
  {NoStop}%
\bibitem [{\citenamefont {Ruiz\char21{}Tijerina}\ \emph
  {et~al.}(2012)\citenamefont {Ruiz\char21{}Tijerina}, \citenamefont
  {Cornaglia}, \citenamefont {Balseiro},\ and\ \citenamefont
  {Ulloa}}]{Tijerina2012}%
  \BibitemOpen
  \bibfield  {author} {\bibinfo {author} {\bibfnamefont {D.~A.}\ \bibnamefont
  {Ruiz\char21{}Tijerina}}, \bibinfo {author} {\bibfnamefont {P.~S.}\
  \bibnamefont {Cornaglia}}, \bibinfo {author} {\bibfnamefont {C.~A.}\
  \bibnamefont {Balseiro}}, \ and\ \bibinfo {author} {\bibfnamefont {S.~E.}\
  \bibnamefont {Ulloa}},\ }\href {\doibase 10.1103/PhysRevB.86.035437}
  {\bibfield  {journal} {\bibinfo  {journal} {Phys. Rev. B}\ }\textbf {\bibinfo
  {volume} {86}},\ \bibinfo {pages} {035437} (\bibinfo {year}
  {2012})}\BibitemShut {NoStop}%
\bibitem [{\citenamefont {Jones}\ and\ \citenamefont
  {Varma}(1987)}]{Jones-Varma1987}%
  \BibitemOpen
  \bibfield  {author} {\bibinfo {author} {\bibfnamefont {B.~A.}\ \bibnamefont
  {Jones}}\ and\ \bibinfo {author} {\bibfnamefont {C.~M.}\ \bibnamefont
  {Varma}},\ }\href {\doibase 10.1103/PhysRevLett.58.843} {\bibfield  {journal}
  {\bibinfo  {journal} {Phys. Rev. Lett.}\ }\textbf {\bibinfo {volume} {58}},\
  \bibinfo {pages} {843} (\bibinfo {year} {1987})}\BibitemShut {NoStop}%
\bibitem [{\citenamefont {Affleck}\ \emph {et~al.}(1995)\citenamefont
  {Affleck}, \citenamefont {Ludwig},\ and\ \citenamefont
  {Jones}}]{Affleck1995}%
  \BibitemOpen
  \bibfield  {author} {\bibinfo {author} {\bibfnamefont {I.}~\bibnamefont
  {Affleck}}, \bibinfo {author} {\bibfnamefont {A.~W.~W.}\ \bibnamefont
  {Ludwig}}, \ and\ \bibinfo {author} {\bibfnamefont {B.~A.}\ \bibnamefont
  {Jones}},\ }\href {\doibase 10.1103/PhysRevB.52.9528} {\bibfield  {journal}
  {\bibinfo  {journal} {Phys. Rev. B}\ }\textbf {\bibinfo {volume} {52}},\
  \bibinfo {pages} {9528} (\bibinfo {year} {1995})}\BibitemShut {NoStop}%
\end{thebibliography}
\end{document}